\documentclass[twocolumn]{aastex63}
\usepackage{amsmath}
\usepackage{amssymb}
\usepackage{graphicx}
\usepackage{color}
\usepackage{float}

\definecolor{darkgreen}{rgb}{0,0.35,0}

\shorttitle{Accretion of Gas Giants}
\shortauthors{Li et al.}

\begin{document}

\title{Accretion of Gas Giants Constrained by the Tidal Barrier}

\author{Ya-Ping Li}
\affiliation{Theoretical Division, Los Alamos National Laboratory, Los Alamos, NM 87545, USA}
\author{Yi-Xian Chen}
\affiliation{Department of Physics, Tsinghua University, Beijing, 100086 China} 
\author{Douglas N. C. Lin}
\affiliation{Department of Astronomy \& Astrophysics, University of California, Santa Cruz, CA 95064, USA}
\affiliation{Institute for Advanced Studies, Tsinghua University, Beijing 100086, China}
\author{Xiaojia Zhang}
\affiliation{Department of Astronomy \& Astrophysics, University of California, Santa Cruz, CA 95064, USA}
\affiliation{Department of Earth Science, Hong Kong University, Hong Kong, China}

\correspondingauthor{Ya-Ping Li, Yi-Xian Chen}
\email{leeyp2009@gmail.com, yx-chen17@mails.tsinghua.edu.cn}
\date{Oct 15, Revised for ApJ}
\begin{abstract}

After protoplanets have acquired sufficient mass to open partial gaps in their natal protostellar disks, residual gas continues to diffuse onto horseshoe streamlines under effect of viscous dissipation, and meander in and out of the planets' Hill sphere. Within the Hill sphere, the horseshoe streamlines intercept gas flow in circumplanetary disks. The host stars' tidal perturbation induces a barrier across the converging streamlines' interface. Viscous transfer of angular momentum across this tidal barrier determines the rate of mass diffusion from the horseshoe streamlines onto the circumplanetary disks, and eventually the accretion rate onto the protoplanets. We carry out a series of numerical simulations to test the influence 
of this tidal barrier on super-thermal planets. In weakly viscous disks, protoplanets' accretion rate steeply decreases with their masses 
above the thermal limit. As their growth timescale exceeds the gas depletion time scale, their masses reach asymptotic values comparable to that of Jupiter. In relatively thick 
and strongly viscous disks, protoplanets' asymptotic masses exceed several times that of Jupiter. { Two
dimensional numerical simulations show} that such massive protoplanets strongly excite the eccentricity 
of nearby horseshoe streamlines, destabilize orderly flow, substantially enhance the diffusion rate 
across the tidal barrier, and elevate their growth rate until their natal disk is severely depleted. 
{ In contrast, eccentric streamlines remain stable in three dimensional simulations.}
Based on the upper fall-off in the observe mass distribution of known exoplanets, we suggest their natal disks had relatively low viscosity ($\alpha \sim 10^{-3}$), modest thickness ($H/R \sim 0.03-0.05$), and limited masses (comparable to that of minimum mass solar nebula model).

\end{abstract}

\section{Introduction}

A widely adopted theory for planet formation is the core accretion scenario \citep{BodenPollack1986,Pollack_etal_1996,Ida_Lin_2004}. In this theory, giant planets form after the emergence of solid cores through accretion of atmosphere in gas-rich protostellar disks (PSDs). During the quasi-steady accretion phase, the growth of the proto-atmosphere around the cores are driven by slow Kelvin-Helmholtz contraction. {When an atmosphere's mass grows beyond the core mass, 
it becomes unstable and the protoplanet transforms into a gas giant through runaway accretion \citep[e.g.][]{mizuno1980,
Pollack_etal_1996,Lissauer2009,PisoYoudin2014,Leeetal2014, LC2015, Ormel2015, Ali-Dibetal2020, chen2020a}.} 

{In the runaway phase, gas inflow onto the core becomes dynamical. The protoplanet's gravity dominates the flow out to the smaller of the 
Bondi radius $R_{\rm B}= G M_{\rm p}/c_{\rm s} ^2$
and the Hill radius $R_{\rm H} = (q/3)^{1/3} a_{\rm p}$. Here $c_{\rm s}$ is the sound speed of disk gas at the planet's semi-major axis $a_{\rm p}$, and $q = M_{\rm p}
/M_\ast$ is the ratio of planet mass $M_{\rm p}$ over stellar mass $M_\ast$.  When the planet mass is
smaller than the thermal limit $q \lesssim q_{\rm th} = 3h_{\rm p}^3$, 
we have $R_{\rm B}\lesssim R_{\rm H}\lesssim H_{\rm p}$, where $H_{\rm p} =c_{\rm s}/\Omega_{\rm p}$ is the disk scale height at $a_p$, $\Omega_{\rm p}$ is the Keplerian angular frequency of the planet, and $h_{\rm p} = H_{\rm p}/a_{\rm p}$ is
the aspect ratio of the disk at $a_{\rm p}$. As more 
gas is accreted, both the Bondi and the Hill radii increase with $q$, although the Hill radius grows less steeply. When 
$q \gtrsim q_{\rm th}$, the hierarchy is reversed to become $R_{\rm B} \gtrsim R_{\rm H}\gtrsim H_{\rm p}$.} 

During their formation, emerging protoplanets tidally interact with their natal disks \citep{GT1980}. \citet{Lin_Papaloizou1986a} 
shows that tidal perturbation of planets with mass ratio exceeding $q_{\rm th}$ is sufficiently strong to induce the formation 
of a gap in the disk near the protoplanets’ orbits and quench its own growth. If the disk depletion timescale is quite short, 
the asymptotic mass of gas giants $M_{\rm p} \simeq M_{\rm th} \equiv q_{\rm th} M_\ast$ near the ice line in PSDs similar to 
the Minimum Mass Solar Nebula (MMSN) are comparable to that of Jupiter \citep{Lin_Papaloizou_1993,Bryden1999}.

More recently, many hydrodynamical simulations show that gas continues to flow across the planet's orbit even 
after the planet reaches the gap-opening mass \citep{Lubowetal1999,Kley2001,Bate2003,DM13,Fungetal2014,DurmannKley2015,Szulagyi2014}. The diffusion of incoming materials across the gap maintains a steady gas profile with a certain 
minimum density. Attempts have been made to analytically and empirically model such steady gap profiles 
\citep{Kanagawaetal2015MNRAS, Duffell2015,  GinzburgSari2018, Duffell2020}, and these gap prescriptions have been 
applied to infer gas giant planets' growth limit in the modified gap-opening paradigm 
\citep{TanigawaIkoma2007, Tanigawa2016,   rosenthal2020consumption}. 
The asymptotic masses obtained in these studies are typically 
an order of magnitude higher than previously estimated, and they predict an excess of $\sim 10\ M_{\rm J}$ planets around
solar-type star systems.  The prolific production of such high-mass planets may be incompatible with the observed 
ceiling in the mass distribution obtained from radial velocity surveys of long-period planets 
\citep{Marcy2005, cumming2008, Mayor2011, petigura2018}. 

In contrast, \citet{Dobbsdixon2007} has shown that despite the residual flow through the gap, 
the gas may not be so efficiently accreted onto the protoplanets.  In general, the protoplanet is surrounded by a circumplanetary disk (CPD) supplied by gas from the PSD via horseshoe streamlines (and vertical flux in 3D cases \citep{Szulagyi2014}). {In an inviscid or weakly viscous disk, there is a vortensity mismatch between the horseshoe streamlines and the 
outer region of CPD.  Under the influence of the host star's tidal perturbation, a tidal barrier is established across the interface where 
flows diverge.  Although turbulence or weak shocks lead to angular momentum transfer and energy dissipation 
along this interface, the mass transfer rate from the horseshoe streamlines to the CPD 
is restrained in the low-viscosity limit.}

In the numerical analysis of \citet{Dobbsdixon2007}, there
is no explicitly specified viscosity other than the inherent numerical viscosity in the computational scheme.  
In this paper, we carry out an extensive series of numerical simulations to verify the validity of their 
tidal barrier conjecture in the low-viscosity limit and and show that it is marginalized in the high-viscosity
limit.  We evaluate the gas giants' accretion rates for a considerable ranges of
relevant model parameters.  In \S \ref{previous}, 
we briefly recapitulate the results from previous investigators. 
In \S \ref{method} we introduce the numerical setup of our simulations, and in \S \ref{results} 
we discuss our numerical results. We explore different parameter spaces of viscosity $\alpha$ and $h_{\rm p}$ to compare our own 
numerical results on accretion rates with the predictions extrapolated from various prescriptions.
We also {{ use gas vortensity contours and velocity vector field of tracer particles}} to show the effectiveness of the 
tidal barrier. In \S \ref{sec:discussion} we discuss about some other outstanding issues, and give a summary of our findings and their implications.

\section{Previous semi-analytic estimates and numerical results}
\label{previous}
In the classical scenario, tidal interaction between a gap-opening planet and the surrounding disk induces 
an exponentially deep gap, enough to disconnect the inner disk and outer disk, and quench the accretion onto the planet. 
However, more simulations showed that there is residual gas in the gap even for planets with 
$q \geq q_{\rm th}$ \citep[e.g.][]{Lubowetal1999,Fungetal2014,DurmannKley2015}. In general, the azimuthally averaged surface density attains a minimum value
$\Sigma_{\rm min}$ at the planet's orbital semi major axis $a_{\rm p}$.  In the two dimensional limit,
a finite residual $\Sigma_{\rm min}$ can supply an embedded planet with an accretion rate of
\begin{equation}
    \dot{m}_{\rm p}=A \Sigma_{\rm min}.
\end{equation}
Based on numerical simulations, \citet{TanigawaWatanabe2002} fitted an empirical formula for the accretion 
coefficient $A$
\begin{equation}
    A=0.29{h_{\rm p}}^{-2}q^{4/3} a_{\rm p}^{2} \Omega_{\rm p}.
\end{equation}

Apart from $A$, we still need an estimate for $\Sigma_{\rm min}$ to calculate $\dot{m}_{\rm p}$. With a global numerical scheme, \citet{DM13} simulated 
tidal interaction between  planets and their natal 
disks for several $M_{\rm p}$, $\alpha$, and $h_{\rm p}$.  They empirically obtained a formula 
for the characteristic gas surface density in the gap region:
\begin{equation}
    \frac{\Sigma_{\rm min}}{\Sigma_{\rm p}}=\frac{1}{1+0.034 K}, \ \ \ \ {\rm where} \ \ \ \ K=q^2h_{\rm p}^{-5}\alpha^{-1}.
    \label{bottomdensity}
\end{equation}

\begin{figure}[htbp]
\centering
\includegraphics[width=0.52\textwidth,clip=true]{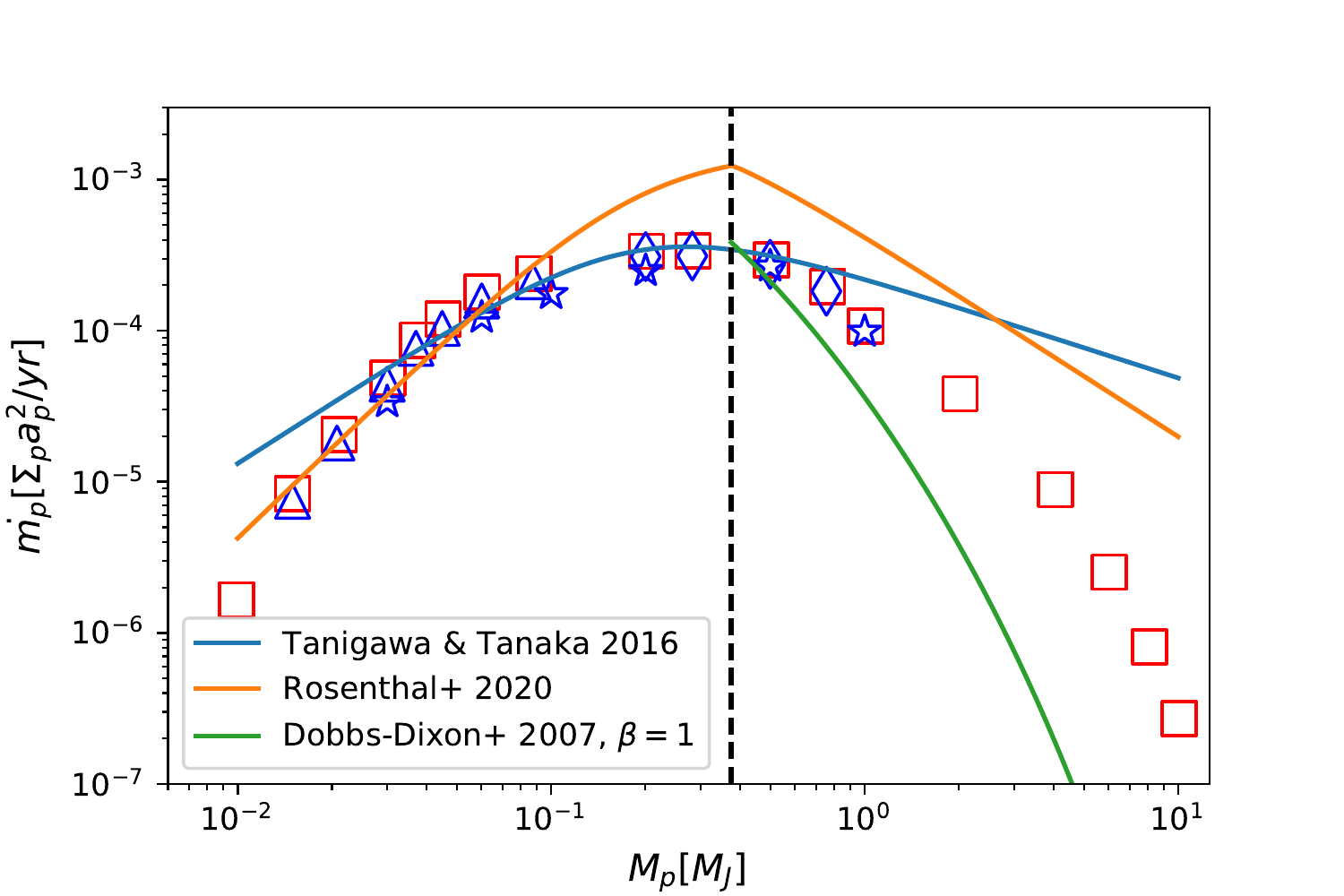}
\caption{Planetary accretion rates as a function of planet mass. Different color curves indication predictions of planet accretion rate, in units of disk mass per year, as functions of planet mass $M_{\rm p}$ for a specific $h_{\rm p}=0.05$ and $\alpha=0.004$. Different blue triangles and stars represent \citet{DAngelo2003}'s data points with different smoothing potential models. The red squares are results from \citet{Boden2013}.}
 \label{fig:predictions}
\end{figure}

This reduction factor in the magnitude of the surface density $\Sigma_{\rm p}$ has also been consistently approximated using semi-analytical approaches by
\citet{Kanagawaetal2015MNRAS}. With this estimate of $\Sigma_{\rm min}$ as well 
as the estimate of $A$ from \citet{TanigawaWatanabe2002}, \citet{Tanigawa2016} obtained a full prescription 
for the planetary accretion rate (hereafter, it is referred to as the TT formula). Their predicted mass 
accretion rate for $\alpha=0.004$ and $h_{\rm p}=0.05$ at $a_{\rm p}=5.2$ au is plotted in Figure \ref{fig:predictions}. {The 
accretion rate of planets with $M_{\rm p} \gtrsim M_{\rm th}$ is constrained by the reduction of $\Sigma_{\rm min}$.
Since the accretion timescale becomes an increasing function of $M_{\rm p}$,  it is no longer a ``runaway" 
process by definition. Nevertheless, on a Myr disk depletion 
time scale, it can lead to asymptotic values of $M_{\rm p} > > M_{\rm J}$. 
For the rest of the paper, we will refer to the accretion of gas giants as ``dynamical" to differentiate it from the thermal contraction process.}

For comparison, we have also shown (in Figure \ref{fig:predictions}) the normalized planetary accretion rates from
independent 3D simulations carried out by \citet{DAngelo2003} (blue, different symbols represent different smoothing potentials) and \citet{Boden2013} (red squares), analogous to Figure 1 of \citet{Tanigawa2016}. There are 
apparently some deviations in the accretion rates obtained from \citet{DAngelo2003,Boden2013} (with exactly the 
parameters $\alpha=0.004$ and $h_{\rm p}=0.05$) compared to the TT formula. In particular, when the planet 
is above the thermal mass (i.e. with $M_{\rm p} \gtrsim M_{\rm th}$), the numerical steady accretion rate follows 
a steeper decline with $M_{\rm p}$ - this discrepancy is much more evident
in comparison with \citet{Boden2013}'s results. 

\citet{Tanigawa2016} suggested the simulated results of \citet{Machida2010} 
might support their prediction that the accretion rate does not fall off steeply with $M_{\rm p}$ above $M_{\rm th}$.
This set of 
simulations used an inviscid disk (with $h_{\rm p}=0.05$ but the physical viscosity $\alpha\approx0$), in which $\Sigma_{\rm min}$
approximated by {Equation} \ref{bottomdensity} is not appropriate. Rather, it might fit into the ``inviscid and sub-thermal" 
regime \citep{GinzburgSari2018}, in which limit $\Sigma_{\rm min}$ is expected to be very small. 
\citet{GinzburgChiang2019} applied the inviscid scaling to obtain the final masses of gas giants and 
found they are generally low ($\lesssim M_{\rm th} \sim M_{\rm J}$), but they also clarified that this scaling suffers from inaccuracies for 
planets with $M_{\rm p} \gtrsim M_{\rm th}$ in disks with small $h_{\rm p}$ (see their \S 3.2.4). We do not attempt to apply this 
scaling to account for \citet{Machida2010}'s results, since we will be focusing on the moderate-viscosity, super-thermal regime 
where the \citet{DM13,Kanagawaetal2015MNRAS} formula is usually practically applied \footnote{We note that the original data 
from Figure 5 of \citet{Machida2010} is rescaled by a factor of 2 when plotted in Figure 1 of \citet{Tanigawa2016}, although 
their vertical axis is both $M_{\rm J}$/yr in a standard MMSN disk ($\Sigma=\sqrt{2\pi}\rho H \approx 140 {\rm g \cdot cm^{-2}}$ 
at 5.2 au, $\rho, \Sigma$ is the gas 3D \& 2D density). The reason for this discrepancy is unclear.}. {The accretion 
rates for this set of disk parameter are also calculated in some earlier simulations \citep{Lubowetal1999,Bate2003}. The accretion rates they obtain are typically higher than the nested-grid simulations.  This difference may be due to low resolution around the central accretion zone (covered by only four grid points) or are the artifacts of numerical viscosity. The earlier results are not shown in Fig \ref{fig:predictions} for simplicity.}

In another semi-analytic approach, \citet{rosenthal2020consumption} used simple Bondi and Hill radii
to approximate the accretion radius of planets with sub-thermal and super-thermal masses respectively. 
We refer to this prescription as the RCGM formula.  Since 
for Bondi accretion $\dot{m}_{\rm p} \sim \Sigma_{\rm min}c_{\rm s}R_{\rm B}^2/H$ \citep{Franketal1992} and for Hill sphere accretion 
$\dot{m}_{\rm p} \sim \Sigma_{\rm min}R_{\rm H}^2\Omega_{\rm p}$, they concluded

\begin{equation}
    A=\left\{\begin{array}{l} A_{\mathrm{Bondi}} = c_1 \dfrac{q^{2}}{h_{\rm p}^{4}} \Omega_{\rm p} a_{\rm p}^{2} \quad \quad \quad
    q\leq q_{\rm th} \\ A_{\mathrm{Hill}}=c_2 q^{2/3} \Omega_{\rm p} a_{\rm p}^{2} \quad \quad \quad q>q_{\rm th}\end{array}\right.
    \label{eq:aprime}
\end{equation}
and fitted the order-unity coefficients $c_1$ and $c_2$ with numerical results.  They assumed
that Hill accretion and Bondi accretion continuously connect at around the thermal mass ratio $q_{\rm th}$. 
\citet{rosenthal2020consumption} eventually {obtained} a result of $c_1=0.5$ and $c_2=2.2$ by calibrating the 
factors according to \citet{DAngelo2003}'s 3D results. Practically, they adopted a formula similar to Equation \ref{bottomdensity} 
as an estimate of $\Sigma_{\rm min}$ for gap-opening planets (see their Equations 24-27). Their scaling of $\dot{m}_{\rm p}$ is also shown in Figure \ref{fig:predictions} for disk viscosity $\alpha=0.004$ and aspect ratio at the planet $h_{\rm p}=0.05$.

In comparison with the TT prescription, RCGM prescription follows a steeper decline with $q$.  Nevertheless,
it over-predicts the accretion rate for gap-opening planets compared to numerical results of 
\citet{DAngelo2003,Boden2013}. In fact, if gap-opening planets accrete at the rates predicted by
either prescriptions, the mass doubling timescale would be $\tau_{\rm p} = M_{\rm p}/{\dot m}_{\rm p} 
\sim 10^4-10^5$ years for a Jupiter-mass planets at $r_{\rm p}=5$ au in a MMSN. If their natal disks deplete over  
a characteristic timescale of $\tau_{\rm dep}\sim 1-3\ {\rm Myrs}$, the asymptotic mass 
of proto gas giant planets would typically be $\gtrsim 10\  M_{\rm J}$.  Such an extrapolation 
would be difficult to reconcile with the apparent ceiling (a sharp decline in numbers for planet mass larger than a few $M_{\rm J}$) of the planetary mass distribution 
\citep{cumming2008, Marcy2005,Mayor2011, petigura2018}. 

{In order to get around this excess mass issue, \citet{Tanaka2020} suggested dynamical
gas accretion occur in advanced stages of protostellar evolution, when photoevaporation of the disk is effective \citep{Owen2011}. However, a shallow power law of $\dot{m}_{\rm p}-M_{\rm p}$ still implies that final mass of planets are sensitive to the mass of the disk when planets form \citep[][see their Figs 5 \& 6]{Tanaka2020}. 
A typical disk mass and photoevaporation rate can reproduce the most frequently seen $\sim 2M_{\rm J}$ planet population, but may not be compatible with the ubiquitous gap structures in PSDs 
\citep{andrews2018, long2018,SeguraCox2020}, which suggest planets may commonly form in different stages of disk clearing.  Small $\tau_{\rm p} (< 1$ Myr) also poses challenges to the 
formation of multiple Jupiter-mass planets around the same host stars, unless they
are coeval within some brief time intervals ($< \tau_{\rm p}$).  }

One way to quench accretion rate is to have more severe gas depletion in the gap. 
But as long as there is sufficient gas flow across the gap, a finite residual surface density 
is still maintained such that the scaling in Equation \ref{bottomdensity}
\citep{DM13,Kanagawaetal2015MNRAS} would be order-unity 
accurate. Severe depletion of the gap can only occur in the limit of $q\gg q_{\rm th}$. 
Under such a condition, the gap-forming planets would undergo classical type II migration
\citep{Lin_Papaloizou_1986}.  The omnipresence of extensive migration would lead to 
an overpopulation of hot Jupiters which is not consistent with the observed period distribution \citep{Ida2013}. The preservation of some residual gas flow 
across the gap may reduce the rate of type II migration and contribute to
retention of cold Jupiters \citep{chen2020b}.

The above analytical or empirical formulae neglect the tidal effects due to 
the non-axisymmetric potential in the proximity of the planets' Hills radius.
In the frame corotating with the planet's orbit, the gas flow is subject to the host stars' tidal perturbation and Coriolis effects. Although its angular momentum varies, the vortensity 
and {Bernoulli constants} of the gas are conserved along streamlines in the inviscid
limit \citep{Korycansky1996}.

The tidal perturbation of low-mass planets (with $q \ll q_{\rm th}$) on the gas flow is generally 
weak near their Hills radius. The planets' gravity overwhelms the gas pressure gradient 
only inside their Bondi radius.  
After the planets have acquired a thermal mass (with $q \gtrsim q_{\rm th}$), their
tidal perturbation and the Coriolis force deflect the horseshoe streamlines of the gas
as it enters into the planets' Hill sphere.  The vortensity of the horseshoe streamlines 
generally does not match that on the outer region of the CPD. Viscous transport of 
angular momentum and dissipation of energy are needed for the gas to cross the interface 
which separates the streamlines in these regions. 


This subtle effect is neglected in the TT and RCGM prescriptions.  Their assumption
that the azimuthal distribution of $\Sigma_{\rm min}$ is uniform implicitly assumes that 
the viscous transport and dissipation have smeared out the horseshoe streamlines in the 
region between $R_{\rm B}$ and $R_{\rm H}$.
\citet{Dobbsdixon2007} emphasize the importance of this effect in the inviscid limit.
They showed that, for Hill accretion conserving the Bernoulli {constant} and vortensity (see their Eqn 28):
\begin{equation}
    A_{\rm Hill} \approx 2\pi R_{\rm H} H_{\rm p}\Omega_{\rm p} \exp\left[-\beta \left (\dfrac{R_{\rm H}}{H_{\rm p}} \right)^2-\dfrac12\right]
\label{eq:abarrier}
\end{equation}
where $\beta$ is an order-unity empirical factor. We refer this approach as the DLL prescription.
This modification from the expression of $A_{\rm Hill}$ in Equation  \ref{eq:aprime} becomes large 
in the limit $R_{\rm H} > H_{\rm p}$ i.e. when the protoplanets' mass is above the thermal limit.  
The important implication of this expression is that the protoplanets' accretion rate decays 
exponentially with some power of $q$. In the inviscid limit, both vortensity and Bernoulli constant are conserved 
along the streamlines \citep{Korycansky1996,Balmforth2001}, flow
from the horseshoe streamlines into the protoplanet’s
Hill sphere and gas accretion onto the CPD requires shock dissipation \citep{Szulagyi2014,Zhu2016,Szulagyi2017}. 
Consequently, only a fraction of the gas passing through this region is retained. The 
inviscid numerical simulations \citep{Dobbsdixon2007} confirmed this barrier. However,
the effect of additional diffusing enabled by intrinsic turbulent viscosity 
(other than implicit numerical viscosity) has yet to be investigated. 

In order to contrast the difference between the three prescriptions, we
have also plotted this prediction (Eqn \ref{eq:abarrier}) in Figure \ref{fig:predictions} (green), 
for the empirical value of $\beta \sim 1$.  The predictions of three accretion rate estimates 
diverge at the high values of $q$, albeit their differences are small for Jupiter-mass planets. 
For $M_{\rm p} >M_{\rm J}$, we 
only have results from \citet{Boden2013} for comparison for disk parameters $h_{\rm p}=0.05$, 
$\alpha=0.004$, whose fall-off slope conforms more with DLL than TT and RCGM prescriptions. The excess accretion rate observed in simulations can be naturally accounted for by non-negligible viscous dissipation.
 
Nevertheless, the numerical simulations shown in Figure \ref{fig:predictions} are performed with viscosity. Viscous stress can lead to angular 
momentum transfer between different streamlines and cause deviations from DLL scaling, which could only serve as lower asymptotic limit of accretion rate. In addition to intrinsic viscosity,
artificial angular momentum transfer may also be introduced by the boundary condition
and spacial resolution in different numerical algorithms. \citet{Dobbsdixon2007}
have pointed out that the 2D local-coordinate (with repeated boundary condition
across an azimuthal patch centered on the planet) simulations of \citet{TanigawaWatanabe2002} 
might not be adequate to capture a gap-opening planet's secular 
perturbation on the global gas flow patterns. For the nested-grid coordinates in the 3D 
simulations of \citet{DAngeloetal2005,Boden2013}, the resolution for gas 
flows at azimuth locations far from the planet is also coarse.  It is possible that 
certain global effects might have been overlooked. In high resolution 2D simulations (with 
uniform grids), a very important finding is that when a planet's $q$ reaches $\gtrsim 
0.003-0.005$, the eccentricity of disk gas streamlines is excited \citep{KleyDirksen2006,Regaly2010,DuffelChiang2015}.  
The radial excursion of streamlines bring external gas well into the gap.  Streamline 
crossing inside the planet's Hill radius also leads to shocks, enhancing vortensity 
diffusion and energy dissipation.  Consequently, the accretion rate of these relatively 
massive planet is elevated. 
In this case, the scaling for stable accretion on circular orbits is no longer appropriate.  
{In this paper, we present the results of global high-resolution simulations. The 
results of our 2D simulations confirm that, over a sufficiently long timescale, eccentricity
excitation by high-mass planets do lead to unstable horseshoe streamlines and consequently
high and variable accretion rates (see discussion in \S\ref{sec:eccentric}).}
{Nevertheless, protoplanets can self-consistently acquire an asymptotic mass of $M_{\rm p} \lesssim 3\ M_{\rm J}$ 
without significantly exciting the eccentricity of the streamlines near their orbits, provided that they form
in regions of their natal disk with relatively small thickness (and therefore a modest thermal mass limit) with modest gas surface density}.

In relatively thick regions of the disk where their thermal mass exceeds the threshold mass for the excitation
of significant streamline eccentricity, their asymptotic mass is reached when the nearby regions of the
disk is severely depleted.  This outcome can lead to even higher terminal masses than those estimated by the TT \& RCGM scaling. 

{However, we note that such extrapolation directly contradicts \citet{Boden2013}'s 
3D nested grid results, which do not show a rise in accretion rate even for $10\ M_{\rm J}$ planets. 
Although the results of \citet{Boden2013} seems to support exponential decay of accretion 
rates for Jupiter-mass protoplanets, the causes (physical or computational) for dichotomy between 
the 2D and 3D simulations for high-mass planets need to be cautiously investigated. Although the simulations 
presented in this paper are mainly 2D, we also present results from 3D simulations.  {Our results also 
indicate that the onset of streamline instability may be mitigated in the 3D limit, which contributes to maintaining a low accretion rate.} We offer 
some discussion on this dichotomy in \S \ref{2d3d}.}


\begin{table*}[htbp]
  \caption{\bf Model parameters and planetary accretion rates}\label{tab:para}
  \begin{tabular}{lcccccc|cc}
     \hline\hline
     model & $q/10^{-3}$ & $\Sigma_{\rm p}$ &  $\alpha$  & $h_{\rm p}$ & $\Delta$ & $f/\tau$ & $\dot{m_{\rm p}}$ & Comments\\
                &  & $(M_{\star}/a_{\rm p}^{2})$ &    &  & ($R_{\rm H}$) & ($\Omega_{\rm p}$) & (${\Sigma_{\rm p}a_{\rm p}^{2}\Omega_{\rm p}}$) & \\
     \hline

     a03h03m025 & 0.25 & $10^{-3}$ &  $1.11\times10^{-3}$ & 0.03 & 0.1 & 5.0 & $7.0\times10^{-6}$ & \\
     \hline
     a02h03m1 & 1.0 &-- &  $1.11\times10^{-2}$ & 0.03 & -- & -- & $5.0\times10^{-5}$ & \\
     
     {a03h03m1} & 1.0 & -- &  $1.11\times10^{-3}$ & 0.03 & -- & -- & {$6.5\times10^{-7}$} & \\
     
      a02h05m1 & 1.0 & -- &  $1.11\times10^{-2}$ & 0.05 & -- &  -- & $2.5\times10^{-4}$ & \\
     
     a03h05m1 & 1.0 & -- &  $1.11\times10^{-3}$ & 0.05 & -- &  -- & $1.0\times10^{-5}$ & \\
     
     {a04h03m1} & 1.0 & -- &  $1.11\times10^{-4}$ & 0.03 & -- & -- & {$6.0\times10^{-8}$} & \\
     
     {a04h05m1} & 1.0 & -- &  $1.11\times10^{-4}$ & 0.05 & -- & -- & {$6.0\times10^{-7}$} & \\
     
     {a03h03m1.mig} & 1.0 & -- &  $1.11\times10^{-3}$ & 0.03 & -- & -- & {\color{red}{$0.7-1.5\times10^{-5}$}} & {mig, eccentric planet} \\
     
      a03h05m1.mig & 1.0 & -- &  $1.11\times10^{-3}$ & 0.05 & -- &  -- & $7.0\times10^{-6}$ & mig\\

     
     \hline
     
           {a02h03m1rs} & 1.0 & -- &  $1.11\times10^{-2}$ & 0.03 & 0.05 & 5.0 &  {$4.5\times10^{-5}$} & convergence \\
     
     
     {a02h03m1ta} & 1.0 & -- &  $1.11\times10^{-2}$ & 0.03 & 0.1 & 2.0 & {$4.5\times10^{-5}$} & tests\\
     
    
    {a02h03m1d} & 1.0 & $10^{-4}$ &  $1.11\times10^{-2}$ & 0.03 & -- & -- & {$5.0\times10^{-5}$} & \\

     \hline
     a02h03m2 & 2.0 & -- &  $1.11\times10^{-2}$ & 0.03 & -- &  -- & $3.0\times10^{-5}$ & \\
     
     {a03h03m2} & 2.0 &-- &  $1.11\times10^{-3}$ & 0.03 & -- & -- & {\color{red}$(0.5-10) \times10^{-7}$} & unstable eccentricity\\

     a02h05m2 & 2.0 & -- &  $1.11\times10^{-2}$ & 0.05 & -- & -- & $1.5\times10^{-4}$ & \\
     
     a03h05m2 & 2.0 & -- &  $1.11\times10^{-3}$ & 0.05 & -- &  -- & $2.0\times10^{-6}$ & \\
     
     \hline
     {a02h03m4} & 4.0 & -- &  $1.11\times10^{-2}$ & 0.03 & -- & -- & {$1.0\times10^{-5}$} & \\
     
     {a03h03m4} & 4.0 & -- &  $1.11\times10^{-3}$ & 0.03 & -- & -- & {\color{red}$(1.0-5.0)\times10^{-5}$} &unstable eccentricity\\
     
     {a02h05m4} & 4.0 & -- &  $1.11\times10^{-2}$ & 0.05 & -- & -- & {\color{black}$8.0\times10^{-5}$} &  \\
     
      {a03h05m4} & 4.0 & --&  $1.11\times10^{-3}$ & 0.05 & -- & -- & {\color{red}$(1.0-3.0)\times10^{-5}$} & unstable eccentricity \\
     
     \hline
     {a02h03m6} & 6.0 & -- &  $1.11\times10^{-2}$ & 0.03 & -- & -- & {\color{red}$(2.0-20.0)\times10^{-5}$} &  unstable eccentricity \\
     {a02h05m6} & 6.0 & -- &  $1.11\times10^{-2}$ & 0.05 & -- & -- & {\color{black}$6.0\times10^{-5}$} &  \\
     \hline
     {a02h05m8} & 8.0 & -- &  $1.11\times10^{-2}$ & 0.05 & -- & -- & {\color{red}$(1.0-2.0)\times10^{-4}$} &  unstable eccentricity \\
     
     \hline
     
     {lowa3d} & 4.0 & -- &  $1.0\times10^{-3}$ & 0.05 & -- & -- & {$7.5\times10^{-7}$} & 3D \\
     {higa3d} & 4.0 & -- &  $4.0\times10^{-3}$ & 0.05 & -- & -- & {$7.0\times10^{-6}$} & 3D \\

     \hline\hline
   \end{tabular}
   
  \end{table*}

\section{Numerical Method}
\label{method}

\subsection{Hydrodymamical Model}
In our simulations with \texttt{LA-COMPASS} \citep{Lietal2005,Lietal2009}, we use a thin and 
non-self-gravitating PSD. We choose a 2D cylindrical coordinate 
system $(r,\varphi)$, and the origin locates at the position of the central star with 
mass $M_{*}$. We adopt a simple model in which the PSD’s temperature is independent 
of the distance above the midplane $T_{\rm disk}\propto r^{-\zeta}$ with aspect ratio
\begin{equation}
    h=\frac{H}{r}=h_{\rm p}\left(\frac{r}{r_{0}}\right)^{(1-\zeta)/2}
\end{equation}

We take the natural units of $G=M_{*}=r_{0}=1$. The planet is fixed at a semi-major axis of $a_{\rm p}=r_{0}$ except for the migrating planet models in \S \ref{sec:mig}. The disk has an initial profile of 
\begin{equation}
    \Sigma(r)=\Sigma_{\rm p}\left(\frac{r}{r_{0}}\right)^{-s},
\end{equation}
where $\Sigma_{\rm p}$ is the surface density at the orbital radius of the planet, with a natural unit of $M_*/a_{\rm p}^2$. 
The 2D velocity vector of the gas is $\vec{v}=\left(v_{{r}}, {v}_{\varphi}\right)$, and angular velocity is $\Omega=v_{\varphi}/r$.

We numerically solve the vertically integrated continuity equation and the equation of motion in a non-rotating frame 
centered on the host star.
In calculating the gravitational potential of the planet at $\vec r$, we use a smoothed potential 
of the form \citep[e.g.][]{GT1980}
\begin{equation}
\phi_{\rm p}=-\frac{G M_{\mathrm{p}}}{(\left|\vec{r}_{\rm p}-\vec{r}\right|^2+\epsilon^2)^{1/2}}+q \Omega_{\rm p}^{2} \vec r_{\rm p} \cdot \vec r
\label{potential}
\end{equation}
where $\vec{r}_{\rm p}$ indicates the location of the planet, $\epsilon=0.4\ R_{\rm H}$ is the softening length.  The second term on the 
right hand side of the above equation corresponds to the indirect term associated with the host star's motion relative to 
the center of mass.  
Without the loss of generality, we can consider a solar-type host star with $M_{*}=M_{\odot}$ and therefore $q=0.001$ corresponds to a planet mass of $M_{\rm p}=M_{\rm J}$.


We use the damping boundary condition \citep{deValborroetal2006} to provide wave killing zones at each radial edge of the disk to prevent wave reflections. 
We adopt the conventional $\alpha$ prescription for 
the kinematic viscosity \citep{ShakuraSunyaev1973}.
In disk regions where distance to the star $r$ is much smaller than the characteristic disk size, 
the radial velocity and disk accretion rate due to viscous diffusion are given 
as \citep[see][for a review]{Franketal1992}:
 \begin{equation}
v_r \sim  \frac{3 \nu}{2 r} 
\end{equation}
\begin{equation}
\dot{M_{\ast}} \sim 3\pi \nu \Sigma = 3\pi \alpha h_{\rm p}^2 a_{\rm p}^2\Sigma_{\rm p} \Omega_{\rm p} 
\cdot \left(\frac{r}{a_{\rm p}}\right)^{1.5-s-\zeta}
\label{eq:mdotstar}
\end{equation}
In a steady state disk without a planet, $\dot{M}_{\ast}$ is independent of $r$.  This state is established with 
$s+\zeta=1.5$. In this study, we fix $s=1.0$ and $\zeta=0.5$ for the unperturbed disk (i.e. the initial 
distribution without the protoplanet's tidal perturbation).  We set the outer boundary 
of the computational domain to be much larger than $a_{\rm p}$ and impose a steady mass influx with 
appropriately extrapolated values of $\Sigma(r)$ and $h(r)$.

In our main runs, we set the fiducial gas surface density $\Sigma_{\rm p}=10^{-3}$. This corresponds to $\Sigma_{\rm p} = 356\ {\rm g \cdot cm}^{-2}$, which is 2.3 times larger than the MMSN value \citep{Hayashi1981}, for $M_*=M_{\odot}$ and $a_{\rm p}\sim 5$ AU. In natural units, the accretion rate is $8.5\times 10^{-9} (h_{\rm p}/0.03)^2 (\alpha/0.001) M_*\Omega_{\rm p}$ and corresponds to $4.8\times 10^{-9} (h_{\rm p}/0.03)^2 (\alpha/0.001) M_\odot\ \mathrm{yr^{-1}}$ for the Jupiter case of $M_*=M_{\odot}$ and $a_{\rm p}\sim 5$ AU. Nevertheless, we note that when the accretion rate is measured in scale-free units $\Sigma_{\rm p}a_{\rm p}^2\Omega_{\rm p}$ or $\Sigma_{\rm p}a_{\rm p}^2/\mathrm{yr}$, any initial $\Sigma_{\rm p}$ will yield similar results as long as the disk has negligible self-gravity, in our case for $h_{\rm p}> 0.03$. 


\subsection{Planetary Accretion}

In order to simulate planetary accretion, we follow \citet{TanigawaWatanabe2002,DAngelo2003,Dobbsdixon2007} and introduce a small sinkhole (with a radius $\Delta$) centered around the planet. We reduce the central surface density $\Sigma_{\rm c}$ of the gas by a factor of $f$ over some time interval $\tau$.
The magnitude of $\Sigma_{\rm c}$ is determined by gas flow patterns at the planetary proximity and it might be quite 
different from the average surface-density minimum ($\Sigma_{\rm min}$) in the gap.  The planet accretes $\approx \pi f\Sigma_{\rm c}\Delta^2$ during this interval of $\tau$.
\citet{TanigawaWatanabe2002} finds that for $\Delta \lesssim 0.1 R_{\rm H}$, the accretion rate converges to a same steady rate 
regardless of what $\tau$ is initially chosen. A steady accretion rate
\begin{equation}
    \dot{m}_{\rm p}=\dfrac{\pi f\Sigma_{\rm c}\Delta^2}{\tau}
\end{equation}
is obtained to mimick the planet's actual gas accretion rate. For smaller $\tau$, the converging timescale would be faster, as would 
the steady gas surface density in the gas removal zone. We adopt a standard sink hole radius of $0.1R_{\rm H}$ and $f/\tau=5
\Omega_{\rm p}$ (meaning the accretion is imposed at each artificial numerical timestep such that we take away $\sim 100\%$
of disk mass inside the sinkhole in a time of $0.2/\Omega_{\rm p}$). A convergence test is presented in \S \ref{sec:convergence}
for different sink hole radii and timescales. 

Table \ref{tab:para} shows the various model parameters for each simulation.  All the numerical
calculations are performed with a logarithmic mesh in the radial direction.  The grid contains $n_r\times n_\varphi=2048 \times 2048$
computational cells.  It covers a disk in radial range from from $r_{\rm min}=0.25\ r_{0}$ to 
$r_{\rm max}=8.0\ r_{0}$. Unless otherwise stated for convergence tests, we use a standard sink hole radius 
and gas removal rate. 
The model parameters for comparison are $h_{\rm p}$, $\alpha$ and $q$.

In most of our simulations, we have fixed the planet's orbital radius. Because our simulation time is much shorter 
than the growth timescales, we also fix the planet mass. Each simulation is equivalent to taking a snapshot of a planet's steady accretion rate for a specified mass \citep{KleyDirksen2006}. We 
focus on the effect of gap formation on the planets' accretion. We discuss how planet migration might 
affect the streamlines within the CPD in \S \ref{sec:mig}. 

{ We have two high-resolution 3D simulations (\S \ref{2d3d}) to test 
the streamline instability as shown in some 2D models. For 3D models, we additionally specify an initial density structure 
in the vertical direction:
\begin{equation}
    \rho_{\rm g}(r,z)=\rho_{0}\left(r/a_{\rm p}\right)^{-\beta_{\rm 3D}}\exp\left({-z^2/2H^2}\right),
\end{equation}
where $\beta_{\rm 3D}=s+(3-\zeta)/2=2.25$ and $\rho_{0}=\Sigma_{\rm p}/\sqrt{2\pi} H_{\rm p}$ are chosen to match the 2D simulation parameters. We maintain the approximation that the disk is locally isothermal and temperature has no dependence on height. We resolve the whole disk with 512 radial grids spaced logarithmically uniform between $r_{\rm min}=0.4\ r_{0}$, $r_{\rm max}=8.0\ r_{0}$, 120 logarithmically uniform grids within 10 disk scale heights from the mid-plane, and 800 uniform grids in azimuth, which is comparable or finer than the central resolution of \citet{DAngelo2003,Boden2013}.}




\section{Simulation Results}
\label{results}

\begin{figure}[htbp]
\centering
\includegraphics[width=0.45\textwidth,clip=true]{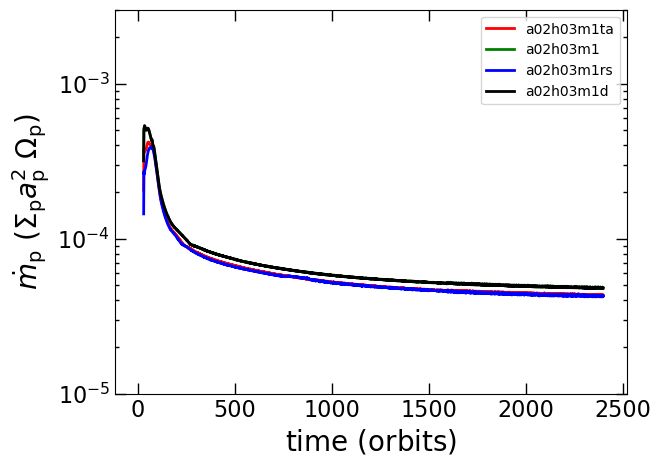}
\caption{The evolution of planetary accretion rate, measured in scale-free unit $\Sigma_{\rm p}a_{\rm p}^2\Omega_{\rm p}$, of the convergence test runs. The model parameters are given in Table \ref{tab:para}.}
 \label{fig:convergence}
\end{figure}

We summarize all of our models in Table \ref{tab:para}. All the 2D models are named according to the disk viscosity $\alpha$, disk aspect ratio $h_{\rm p}$ and planet mass ratio $q$ (e.g., in model a02h03m1, $\alpha=1.11\times10^{-2}$, $h_{0}=0.03$, $q=10^{-3}$). 

\subsection{Convergence test for the accretion rate}
\label{sec:convergence}

In Figure \ref{fig:convergence} we present the evolution of planetary accretion rates for the \texttt{a02h03m1} model with $\alpha=1.11\times 10^{-2}, h_{\rm p}=0.03$, $q=10^{-3}$.  For convergence tests, we also
introduce three variants using the same model parameters but different 
sink hole radius, removal rate, and initial surface density. Case \texttt{a02h03m1rs} has $\Delta=0.05R_{\rm H}$,
\texttt{a02h03m1d} has $\Sigma_{\rm p} = 10^{-4}M_*/a_{\rm p}^2$ and \texttt{a02h03m1ta} has $f/\tau=2\Omega_{\rm p}$ respectively.  Their accretion rates are measured in scale-free unit $\Sigma_{\rm p} 
a_{\rm p}^2\Omega_{\rm p}$ and they are shown in the $\dot{m}_{\rm p}$ column of Table \ref{tab:para}. All
four cases yield very similar steady dimensionless accretion rate, and demonstrate convergence for a) different sink hole radius as long 
as $\Delta\leq 0.1R_{\rm H}$; b) different $f/\tau$ and c) different $\Sigma_{\rm p}$. Indeed a) and b) are already confirmed in detail 
by \citet{TanigawaWatanabe2002}, and c) is trivial as long as disk self-gravity is not invoked.

\subsection{Dynamical growth with large disk eccentricities}
\label{sec:eccentric}

\begin{figure}[htbp]
\centering
\includegraphics[width=0.45\textwidth,clip=true]{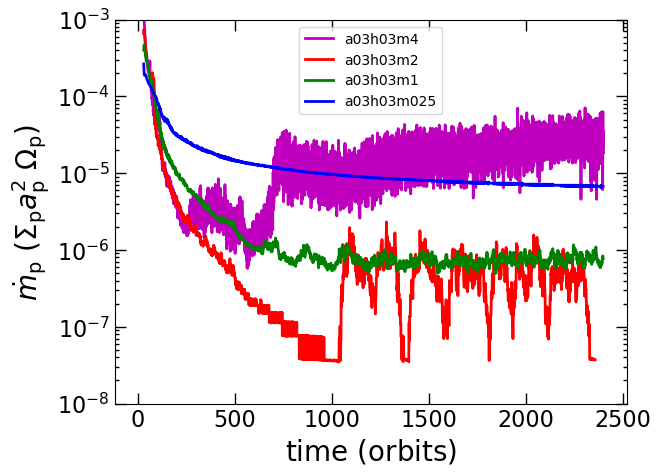}
\includegraphics[width=0.45\textwidth,clip=true]{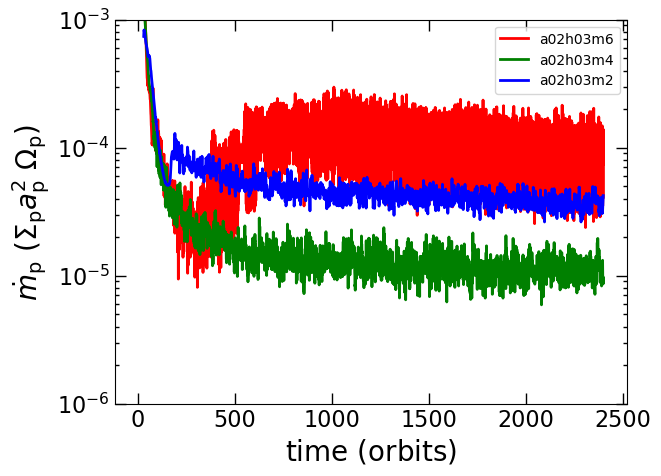}
\caption{The evolution of planetary accretion rate, measured in scale-free units, of cases with $h_{\rm p}=0.03$ and $\alpha=1.11\times 10^{-3}$ (upper panel), and with $h_{\rm p}=0.03$ and $\alpha=1.11\times 10^{-2}$ (lower panel). The planet mass ratios for the upper panel, in increasing order, are $(0.25, 1, 2, 4)\times 10^{-3}$. The lower panel shows the planet mass ratios of $(2, 4, 6)\times 10^{-3}$. The last two cases in the upper panel and the last case in the lower panel show sudden rise and strong fluctuations of planetary accretion rates. }
 \label{fig:eccentric}
\end{figure}

\begin{figure*}[htbp]
\centering
\includegraphics[width=0.8\textwidth,clip=true]{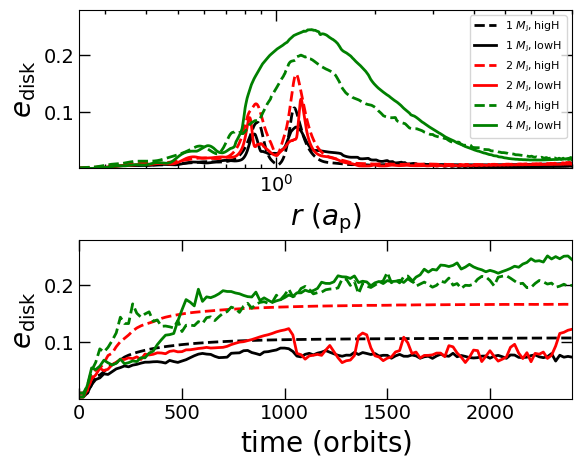}
\caption{Top panel: The azimuthally averaged radial distribution of disk eccentricity at t=2400 orbits. Lower panel: the growth of the \textit{peak} eccentricity within the disk over time. The solid lines represent the low disk scale height $h_{\rm p}=0.03$ cases, and the dashed lines represent the high scale height $h_{\rm p}=0.05$ cases.}
 \label{fig:eccentricevolution}
\end{figure*}

We list the numerical results of the accretion rates for all the models in the $\dot{m_{\rm p}}$ columns of Table \ref{tab:para}. 
Similar to the convergence tests, most of the cases have a steady asymptotic accretion rate. However, some of the low 
viscosity models generate {significant} growth and fluctuation of accretion rate, 
analogous to the findings of \citet{KleyDirksen2006}.

We plot in the upper panel of Figure \ref{fig:eccentric} the evolution of planetary accretion rates for the \texttt{a03h03m025}, \texttt{a03h03m1}, \texttt{a03h03m2}, \texttt{a03h03m4} cases 
with same viscosity/scale height ($[\alpha,h_{\rm p}]$=$[1.11\times10^{-3}$, $0.03]$) but increasing planet mass (equivalently $q$). 
For $M_{\rm p}\leq M_{\rm J}$, the planetary accretion rates 
evolve into steady values after more than 2400 orbits. However in the case of $2\ M_{\rm J},4\ M_{\rm J}$, the accretion 
rates decay rapidly in the first 1000 orbits.  They then abruptly jump to much higher and more unstable values. Similar trend 
appears for the $4\ M_{\rm J}$ case with $h_{\rm p}=0.05, \alpha=1.11\times 10^{-3}$. In the lower panel of Figure \ref{fig:eccentric}, we show the evolution of planetary accretion rates for $2\ M_{\rm J},4\ M_{\rm J}$ and $6\ M_{\rm J}$ with $\alpha=1.11\times10^{-2}$ and $h_{\rm p}=0.03$. This accretion rate fluctuation becomes more prominent at a higher planet mass ($\gtrsim6\ M_{\rm J}$). This pattern is directly associated with the excitation of 
streamlines' eccentricity and it should be distinguished from the orderly accreting process for lower-mass planets. The range 
for accretion fluctuation with the associated streamlines' eccentricity is highlighted in red in Table \ref{tab:para} for these cases. 

Quantitatively, we compute the radial distribution of 
streamlines' eccentricity with the method introduced by \citet{Ogilvie2001} \citep[also see][]{KleyDirksen2006,
TeysOgilve2017}. We follow the evolution of the peak eccentricities for streamlines in the vicinity
of the planet's orbit for each case. Figure \ref{fig:eccentricevolution} shows the eccentricity 
evolution of most of the low-viscosity ($\alpha=1.11\times10^{-3}$) cases. 
The top panel shows the radial eccentricity distribution of disk gas at 2400 orbits, and the bottom panel shows the growth of the maximum eccentricity value within the disk in 2400 orbits. The disk eccentricity for a certain radius $r$ is given by azimuthally averaging the eccentricity for every cell at this radius, assuming the fluid elements are on Kepler orbits around the central star. 

In this regime, we note that for a planet mass of $M_{\rm p}=M_{\rm J}$ in the $h_{\rm p}=0.03$ case, and for $M_{\rm p}= 1$ \& $2M_{\rm J}$ in 
the $h_{\rm p}=0.05$ case, the peak eccentricities converge to a steady and stable value after $\sim2000$ orbits.
In the $M_{\rm p} = 2$ and $4 M_{\rm J}$ with $h_{\rm p}=0.03$ cases and the $M_{\rm p}= 4 M_{\rm J}$ with 
$h_{\rm p}=0.05$ case, the streamline eccentricity grows to be $\gtrsim R_H/a_{\rm p}$ and the streamlines near the 
gap edge become unstable.  In these cases, the planet's accretion rate does not converge to a steady value
even after a few thousand orbits. The fluctuating eccentricities allow much more 
efficient angular momentum transfer and energy dissipation between crossing streamlines.  They enhance the replenishment of materials 
into the planets' Hill radius during one dynamical time.  These fluctuating cases cannot be accurately predicted by the existing analytic approximation of 
orderly-accretion scenarios. Regarding the tidal barrier, large fluctuation in eccentricity also breaks the time-independence 
of the flow structure, therefore the constraint inferred from the assumption of Bernoulli constant and vortensity conservation 
would be loosened.

For the high viscosity ($\alpha=1.11\times 10^{-2}$) cases, we confirm the results of \citet{KleyDirksen2006} that
for $M_{\rm p} \lesssim 6 M_{\rm J}$, 
the horseshoe streamlines are stable, albeit small eccentricities are excited.  They suggest that for $\alpha=0.01, 
h_{\rm p}=0.05$, a mass of $\gtrsim 5\ M_{\rm J}$ is needed to excite sufficient eccentricity to ``stir up" accretion
whereas we found similar results for $M_{\rm p} \gtrsim 6 M_{\rm J}$. For the $h_p=0.03$ case, this threshold 
is lower and for $M_{\rm p} \gtrsim 4 M_{\rm J}$ we already see the rise in planet accretion rate. 
{Generally, low viscosity and low disk scale height favors eccentricity excitation \citep{zhang2018}. 
However, our 2D results are in disagreement
with the results of 3D nested-grid simulations by \citet{Boden2013}.  Their simulations do not indicate the onset of instability 
due to streamline eccentricity even in the high $M_{\rm p}$ line, for a medium viscosity $\alpha=4\times 10^{-3}$ that lies between our high and low cases. Consequently, the tidal barrier limit is always effective and
their results matches with that predicted by the DLL prescription. { We focus on results from our 2D high resolution runs in the main content of this paper, but in \S \ref{2d3d} we also report this dichotomy between the 2D and 3D simulations. }}

\subsection{Tidal Barrier for Jupiter-size Giants: Accretion Rates and Flow Patterns}

After distinguishing the cases with unstable streamline eccentricities and risen accretion rates, we compare the numerical results with 
predictions inferred from various prescriptions.  This scrutiny is particularly relevant for models with orderly 
horseshoe streamlines and steady ${\dot m}_{\rm p}$ which are the main focus of this paper. To show all of our results in the 
manner of Figure \ref{fig:predictions}, we convert our dimensionless accretion rates into units of disk mass 
($\Sigma_{\rm p} a_{\rm p}^2$) per \textit{Earth year}, placing the planet on a fixed circular orbit with a semi major axis $a_{\rm p}=5$ AU.

\begin{figure*}[htbp]
\centering
\includegraphics[width=1\textwidth,clip=true]{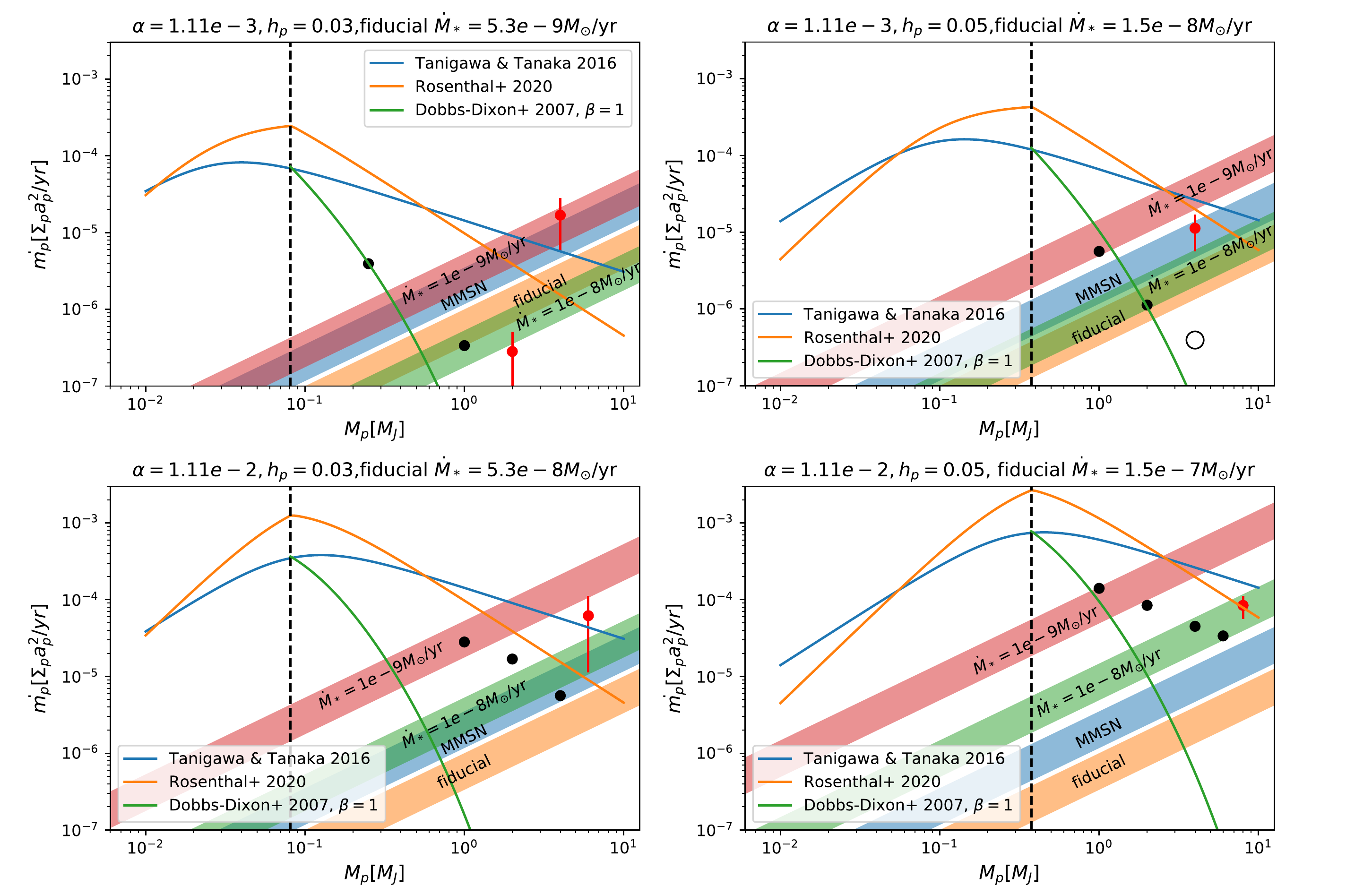}
\caption{Planetary accretion rates as a function of planet mass in normalized units. Different 
color curves indication predictions of planet accretion rate as a function of planet mass $M_{\rm p}$, for four sets of $h_{\rm p}$ and $\alpha$ value. The black dots represent our numerical results for planet mass accretion rate. We add error-bars for the cases with unstable accretion rate and eccentricity. Color bands indicate where the growth timescale of the planet falls between 1-3 Myrs, for four different specific values for $\Sigma_{\rm p}$/$\dot{M}_{\ast}$.{ On the upper right panel we also plot the result from our 3D case with $M_{\rm p}=4 M_{\rm J}$, $h_{\rm p}=0.05$ and $\alpha=10^{-3}\approx 1.11\times 10^{-3}$ as a black open circle, relevant discussion see \S \ref{2d3d}}.}
\label{summary}
\end{figure*}

Four sets of different $\alpha$ and $h_{\rm p}$ are used in our simulations.   The final accretion rates 
${\dot m}_{\rm p}$ from these simulations are plotted (in Figure \ref{summary}) with the inferred values of 
${\dot m}_{\rm p}$ for the three prescriptions as a function of $M_{\rm p}$ (in units of $M_{\rm J}$). 
Since the results of additional convergence tests are similar to those of the standard case, they are 
not shown here. For cases with large long-term fluctuations in the planets' accretion rates, 
we plot the medium value with solid red points and indicate the upper and lower bound with error-bars. 

Results from the simulation agree well with the DLL tidal-barrier prescription in the low-viscosity 
($\alpha = 1.11\times 10^{-3}$) limit as it was intended.  For relatively large $\alpha ( =1.11\times 10^{-2}$),
viscous stress can lead to angular momentum transfer and energy dissipation across the
tidal barrier.  As gas distribution is smoothed over the gap region and inside 
the Hill radius (which is implicitly assumed by the RCGM prescription), 
the enhanced gas supply to the outer regions
of the CPD elevates ${\dot m}_{\rm p}$ onto the planets to values between those predicted by 
different prescriptions.  Such high ${\dot m}_{\rm p}$ may enable $M_{\rm p}$ to cross the threshold mass
for the excitation of unstable streamline eccentricity, rapid diffusion of gas across 
the tidal barrier, and inflated ${\dot m}_{\rm p}$.  The growth of these planets' $M_{\rm p}$  (red dots) 
continues, until their natal PSD's are severely depleted with a $\Sigma_{\rm p}$ less than that of MMSN 
, or $\dot{M}_{\ast} < 10^{-8} M_\odot$ yr$^{-1}$.

The conversion of ${\dot m}_{\rm p}$ in Figure \ref{summary} into physical units requires the specification of 
$\Sigma_{\rm p}$.  The actual value of $\Sigma_{\rm p}$ in physical units can be inferred from $\alpha$, $h_{\rm p}$, and the mass accretion rate $\dot{M}_\ast$ in the PSD (Equation  \ref{eq:mdotstar}).  These values in physical 
units enable the determination of the characteristic growth time scale, $\tau_{\rm p} = M_{\rm p}/{\dot m}_{\rm p}$ for 
a protoplanet with a given mass $M_{\rm p}$.  Inversely, we can infer the matching $\dot{M_{*}}$ or
$\Sigma_{\rm p}$ required for any planet to acquire, but not to over-ingest, an asymptotic 
mass $M_{\rm p}$ within the typical PSD's depletion timescale ($\tau_{\rm dep} \simeq 2 \pm 1$ Myr) 
\citep{Hartmannetal1998}).

We use four color bands to show this inversion (Figure \ref{summary}), representing the range 1 Myr $<\tau_{\rm p}<$ 3 Myr. 
The orange color band represents
a fiducial model in which we set $\Sigma_{\rm p} = 356$ g cm$^{-2}$, such that the characteristic disk mass $\Sigma_{\rm p} a_{\rm p}^2 = M_{\rm J}$ (used as the standard in simulations). In comparison, the blue band is analogous to the less massive
MMSN model \citep{Hayashi1981}. {In the standard MMSN, $\Sigma_{\rm p,0} = 143$ g cm$^{-1}$ at $5.2$AU, Jupiter's 
present-day semi major axis. In the dynamical accretion context, we adopt a modified MMSN model with $\Sigma_{\rm p} = 100$ 
g cm$^{-2}$ at planet location. This value corresponds to extracting a surface density from its 
initial value of $\Sigma_{\rm p, 0}$ } 
{ to $t=1$ Myr over a characteristic depletion timescale of $\tau_{\rm dep}=3\ {\rm Myrs}$,
\begin{equation}
    \Sigma_{\rm p} (t)=\Sigma_{\rm p,0}\exp{(-t/\tau_{\rm dep})}.
\end{equation}
}

We also use pink, and green color bands to represent PSDs with disk accretion rates 
(Equation  \ref{eq:mdotstar}), $\dot{M}_{\ast} =10^{-9}$ and $10^{-8} M_\odot$ yr$^{-1}$ respectively.   The value of
$\dot{M}_{\ast}$ can be inferred directly from observation.


For each set of $\alpha$ and $h_{\rm p}$, the color ${\dot m}_{\rm p}-M_{\rm p}$ bands intercept the three different 
formulae (Figure \ref{summary}).  \citet{Dobbsdixon2007}'s tidal barrier prescription, if applicable, can account for
emergence of planets with asymptotic mass comparable to that of Jupiter from natal disks with
$\dot{M}_{*}$ comparable to that observed in typical disks around classical T Tauri stars.  However, 
the TT and RCGM prescriptions predict much larger (by a factor of $\gtrsim 5$) asymptotic mass for 
emerging planets from identical PSDs.  Equivalently, for the same growth limit, the PSDs need to be more 
severely depleted (with an order of magnitude smaller $\dot{M}$) in the TT, and RCGM prescriptions
than the DLL prescription.

{With a low viscosity ($\alpha= 1.11\times 10^{-3}$), the planetary accretion rates of super-thermal planets 
${\dot m}_{\rm p}$ are considerably lower than predicted by the TT or RCGM prescriptions
either for $h_{\rm p}=0.03$ or $h_{\rm p}=0.05$. The $h_{\rm p}=0.03$ case gives the lowest accretion rate in our
simulations. For the fiducial (orange) or MMSN (blue) models, the mass-doubling timescale 
$\tau_{\rm p} = M_{\rm p} / {\dot m}_{\rm p} \sim 5$ Myrs for a $M_{\rm p} = M_{\rm J}$ protoplanet.  
Since $\tau_{\rm p} \gtrsim \tau_{\rm dep}$,
the gas supply from the PSD would be depleted before the protoplanet can acquire sufficient mass to 
excite unstable streamline eccentricities.  For the $h_{\rm p}=0.05$ case which is more realistic for the MMSN
\citep[e.g.][]{GaraudLin2007}, $\tau_{\rm p} > 1$ Myr for $M_{\rm p}=2 M_{\rm J}$ planets in a fiducial or MMSN model. It is also
possible that planets can only acquire modest masses prior to disk depletion in such environments, and unstable streamline eccentricity would not be excited. }

{With relatively large ($\alpha = 1.11 \times 10^{-2}$) viscosity (lower panels of Figure \ref{summary}, also see 
\S \ref{flowstructure}), however, the tidal barrier is overrun. Viscous diffusion leads to a greater 
flux of gas supply from 1) the outer PSD onto the horseshoe streamlines, 2) horseshoe streamlines at the 
edge of the gap towards the interior of the gap near the $L_4$ and $L_5$ points, and 3) most importantly, 
horseshoe streamlines to the outer region of the CPD. These flow patterns produce the boundary conditions 
assumed by the TT and RCGM prescriptions such that the values of $\dot{m}_{\rm p}$ from the numerical simulations 
come towards their predicted values.}

\begin{figure}[htbp]
\centering
\includegraphics[width=0.48\textwidth,clip=true]{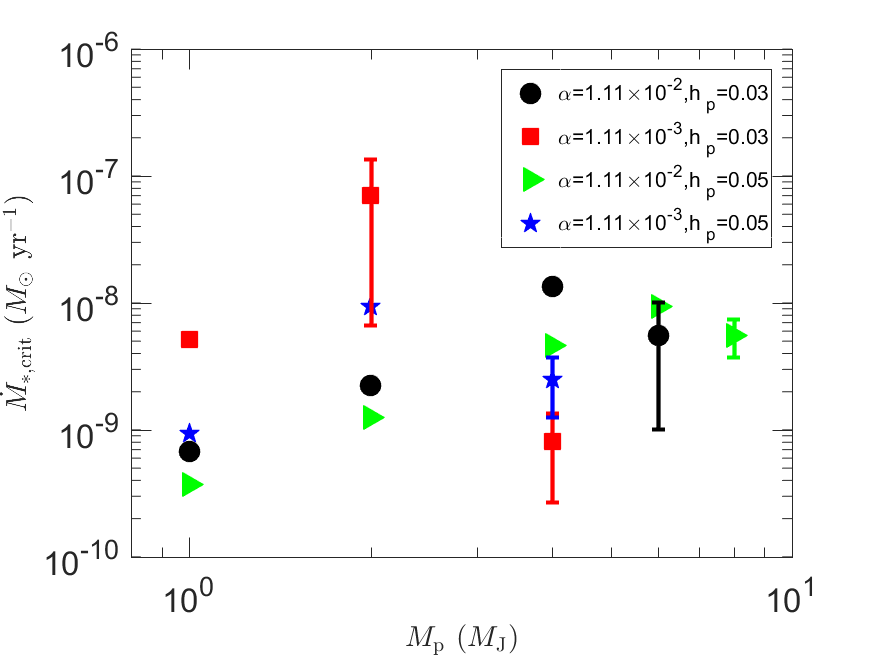}
\caption{The required PSD accretion rate for planet mass doubling as a function of planet mass. The PSD accretion rate in this figure is estimated with a fixed depletion timescale of $3$ Myr. The data points with error bars correspond to the cases where unstable accretions are involved.}
 \label{fig:mdotstar}
\end{figure}

An implication of the numerical results is well illustrated by the upper limit of ${\dot M}_\ast$  that can \textit{prevent} 
planets from doubling their $M_{\rm p}$ over 3 Myr in a steady state disk, shown in Figure~\ref{fig:mdotstar}, denoted as ${\dot M}_{\rm \ast,crit}$. 
Various values of the model parameters $\alpha$ and $h_{\rm p}$ are denoted by different symbols. The low-viscosity 
($\alpha = 1.11\times 10^{-3}$) cases show a peak $\dot{M}_{\rm \ast,crit} (\sim 10^{-8}-10^{-7} M_\odot\ {\rm yr}^{-1})$ around 2 $M_{\rm J}$.
For the high-viscosity 
($\alpha = 1.11\times 10^{-2}$) cases, this peak $\dot{M}_{\rm \ast,crit} (\sim 10^{-8} M_\odot {\rm yr}^{-1})$ is around 4-6 $M_{\rm J}$.
Since this range of ${\dot M}_{\rm \ast,crit}$ is comparable to that observed among PSDs around a few Myr old classical 
T Tauri stars, {the dynamical growth of protoplanets may indeed be halted with masses comparable to Jupiter.}  In disks with
larger $\dot{M_{\ast}} (> {\dot M}_{\rm \ast,crit}$) or $\tau_{\rm dep} (\gtrsim 3 {\rm Myr})$, $M_{\rm p}$ may continue to increase until 
unstable streamline eccentricity is excited.  Such massive planets (with $M_{\rm p} > 4M_{\rm J}$) have smaller growth-quenching
$\dot{M}_{\rm \ast,crit} (\lesssim 10^{-9} M_\odot {\rm yr}^{-1})$ and their $M_{\rm p}$ continues to increase until their natal disk 
is globally depleted by either planetary consumption or outflow. (Local gas depletion in the gap may not be adequate 
to quench planet's growth due to gas supply from other regions of the disk).

{In strongly viscous PSDs around classical T Tauri stars (with ${\dot M_{\ast}} \sim 10^{-8} M_\odot$ yr$^{-1}$), 
$M_{\rm p}$ can reach $\sim 4 M_{\rm J}$ before the excitation of unstable streamline eccentricity (in contrast 
to the low-viscosity cases).  But, planets with $M_{\rm p} (\gtrsim 5-8 M_{\rm J}$) still excite unstable streamline 
eccentricity as we summarize in Table\ \ref{tab:para} (also see \citealt{KleyDirksen2006}).  Transition to unstable streamline 
eccentricity further enhance the planets' accretion rate ${\dot m}_{\rm p}$, reduce the growth-quenching 
${\dot M}_{\ast,{\rm crit}}$, and promote planets' asymptotic masses to become much larger than that of Jupiter.  Nevertheless, the growth of planets can be halted at $M_{\rm p} \sim 1-2 M_{\rm J}$ with ${\dot M}_\ast (\sim 10^{-9} M_\odot {\rm yr}^{-1})$ comparable to or smaller than that found in PSDs 
during some advanced evolutionary stages (Figure \ref{fig:mdotstar}), including 1) those around weak-line T Tauri stars or
2) in transitional disks during the brief epochs of rapid gas depletion. Such high planetary accretion rates for the high-viscosity
numerical models are in better agreement with the TT or RCGM scaling laws than the low-viscosity numerical models.  However, 
they also render the asymptotic mass of giants to depend sensitively on the disk mass and stellar accretion rate \citep{Tanaka2020}.
These high-viscosity models introduce some challenges in the modeling of the observed mass function of exoplanets
especially around stars with multiple gas giant if these planets are formed over a range of time prior to or during  
disk clearing stage (see discussions in \S \ref{previous}). They are also difficult to reconcile with the simulations 
\citep{zhang2018, lodato2019} of the ALMA maps of PSDs with gaps
\citep{andrews2018, long2018} and the sparsity of massive gas giants with wide orbits \citep{meshkat2017, 
vigan2017, bowler2018}.  Based on these considerations, we suggest 
typical Jupiter mass giants were born in disks with relatively low viscosity.
This inference is consistent with numerical simulations of magneto-rotational instabilities
in PSDs \citep{bai2013}.  It is also consistent with the models of gap profile based
on ALMA observations \citep{rafikov2017, ansdell2018} albeit the effects of eccentric streamline
instability should be included in such models. This instability modifies that surface density
distribution and flow pattern (Fig. \ref{fig:eccentricevolution}) and may introduce uncertainties
in the $\alpha$ estimates based on these models.}

\subsection{Flow near planet's Hill sphere}

\label{flowstructure}
\begin{figure}[htbp]
\centering
\includegraphics[width=0.5\textwidth,clip=true]{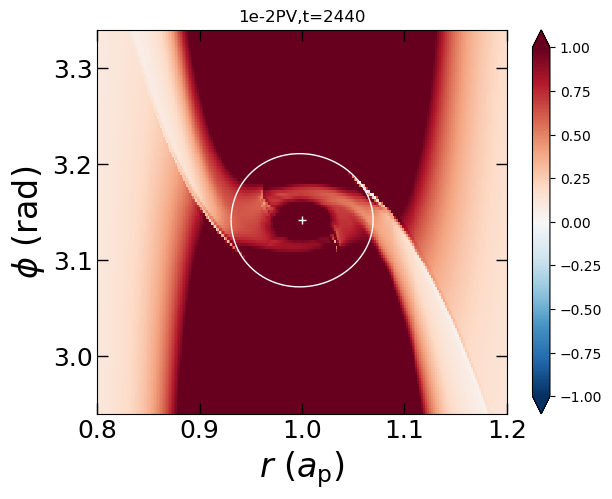}
\includegraphics[width=0.5\textwidth,clip=true]{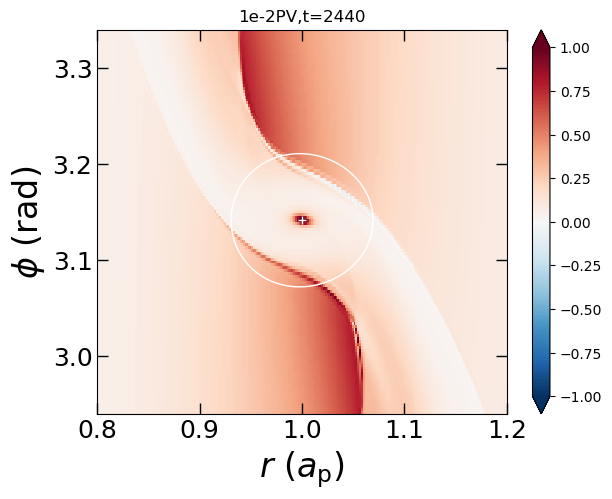}
\caption{The vortensity $\varpi$ distribution around the location of the planet. Upper panel: $M_{\rm p}=M_{\rm J}$, $h_{\rm p}=0.05$, low viscosity case; Lower panel: $M_{\rm p}=M_{\rm J}$, $h_{\rm p}=0.05$, high viscosity case. The circle  around the planet has a radius of $R_{\rm H}$. 1e-2PV means that the value of vortensity shown in this plot is 1\% of the actual value (in dimensionless units) for the purpose of illustration. }
 \label{vortensity}
\end{figure}

The presence of a tidal barrier is best illustrated with velocity-arrow plots and with vortensity contours in the
planetary proximity. The vortensity (or potential vorticity; PV) is defined as \citep{Papaloizou1989,Korycansky1996}:
\begin{equation}
    \varpi=\frac{\omega+2 \Omega_{\rm p}}{\Sigma},
\end{equation}
where $\omega$ is the vertical component of the vorticity, which is defined as
\begin{equation}
\omega = \hat{z}\cdot\nabla\times \vec{v}.
\end{equation}
In the absence of shock dissipation and viscous diffusion, the vortensity is conserved along streamlines.
Moreover, there is generally a mismatch of vortensity between the flow in the outer boundary of the CPD and the horseshoe 
streamlines that are deflected away from the CPD even though their interface is inside the Hill radius
\citep{Korycansky1996}. 

In Figure \ref{vortensity} we compare the vortensity around the planet for case \texttt{a03h05m1} and \texttt{a02h05m1} after 2440 orbits. 
In the first case there is a low viscosity ($\alpha = 1.11 \times 10^{-3}$, upper panel) such that the vortensity is not strictly but 
approximately conserved along streamlines.  Sharp transitions in the vortensity distribution across the CPD-horseshoe
streamline interface is evident. However, for large viscosity ($\alpha=1.11\times10^{-2}$, lower panel) case, viscous 
transport and energy dissipation smooth the distribution of vortensity across these interfaces. Accordingly, the 
tidal barrier for the diffusion of gas in horseshoe streamlines onto the outer regions of the CPD is partially unblocked. 

In order to show the flow pattern of materials across the planet's orbit, we uniformly insert a thin layer of very small 
($2\mu$m) tracer particles outside (at $r>1.2a_{\rm p}$) the gap region.  They are released at t=2400 orbits after the disk has 
attained an asymptotic steady gap structure.  These particles are well coupled to the gas and do not have any feedback 
effect on the disk gas, similar to the scalar dye used by \citet{Duffel_etal_2014}.  They are employed as a passive contaminant 
to characterize the flow pattern. We follow the surface-density and velocity evolution of these particles as they flow into 
and across the gap region.  

\begin{figure}[htbp]
\centering
\includegraphics[width=0.49\textwidth,clip=true]{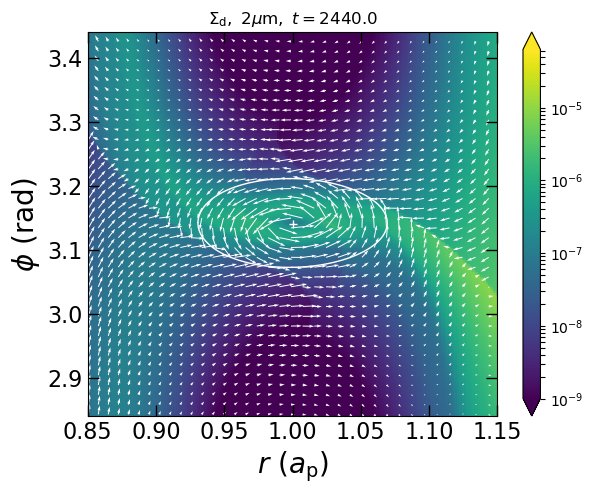}
\includegraphics[width=0.49\textwidth,clip=true]{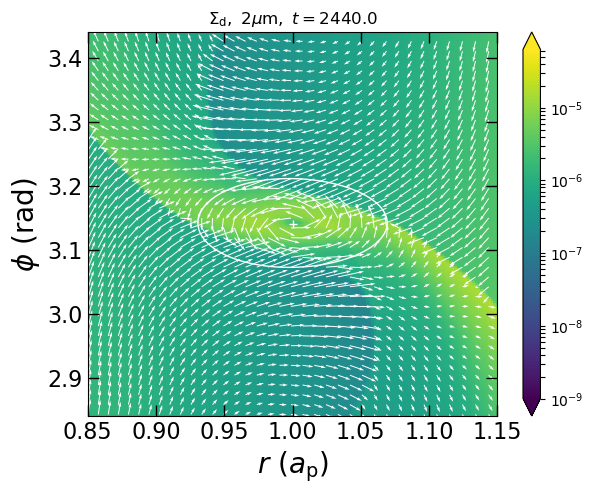}
\caption{The surface density contour and the velocity field in the rotating frame of the planet for  tracer particles. The color represents the surface density of $2\mu$m tracer particles at t=2440 orbits. The particles are uniformly released outside $1.2 a_{\rm p}$ after the gap has formed, and 
completely couple with the gas flowing into the gap region. The arrows show the velocity field of the tracer. Upper panel: $M_{\rm p}=M_{\rm J}$, $h_{\rm p}=0.05$, low viscosity case; Lower panel: $M_{\rm p}=M_{\rm J}$, $h_{\rm p}=0.05$, high viscosity case. The circle around the planet has a radius of $R_{\rm H}$.}
 \label{streamlines}
\end{figure}

Figure \ref{streamlines} shows the surface density $\Sigma_{\rm d}$ and velocity field of these tracer particles, at 40 planetary orbital periods after their initial installment.  The upper and lower panels are for the low and high viscosity cases (\texttt{a03h05m1} and \texttt{a02h05m1} respectively).  In both models, $h_{\rm p}=0.05$. In the low-viscosity case, the tracer particles from the outer 
disk diffuse onto a narrow strand of horseshoe streamlines analogous to the solar system asteroids which share Jupiter's orbit.
These tracer particles do not reach the interior region of the gap (at $r\simeq a_{\rm p}$ and in azimuth away from the planet).
As the streamlines approach the planet, their courses are deflected by the planet's gravitational perturbation and the Coriolis force associated with the rotating frame.  Their paths first diverge outside the planet's Hill sphere along the surface of vortensity discontinuity, between the trailing density waves and the narrowly-confined horseshoe streamlines around the $L_4$ and $L_5$ points in the Roche equipotential surfaces.\footnote{In isolation, local viscous diffusion and energy dissipation around these potential maxima would lead to gas flowing away from them \citep{Lin1987} unless 
there are additional viscous stress (exerted by PSD beyond the gap) or external tidal torque (due to 
other planets) to fill in the gap region. While Figure \ref{streamlines} only show the velocity of incoming tracers at a narrow azimuthal region, horseshoe flows of residual gas circulating $L_4$ \& $L_5$ points can be directly seen in Figure 9 of \citet{chen2020b}.} 

Gas on the horseshoe streamlines flows across the gap, diffuses into 
regions of the PSD interior to $a_{\rm p}$, and preserves a steady ${\dot M}$ towards the central star. This 
flow pattern enable the gas to maintain a steady $\Sigma$ distribution. The associated torque density determines 
the pace and direction of the planet's type II migration \citep{chen2020b}.

Inside the protoplanet's Hill sphere, the horseshoe streamlines converge with those near the outer boundary of the CPD.  
There are also vortensity discontinuity and shear across their interface.  Viscous transfer of angular momentum (relative 
to the protoplanet) enables a small fraction of the intercepting horseshoe streamlines to join onto the CPD.  This fresh 
supply of gas is diffused to the surface of the planet through the viscous transfer of angular momentum in the CPD.  
In the low-viscosity limit, the mass flux through these region is relatively low.  

In the strongly-viscous case, the high efficiency of angular momentum transfer leads to mass diffusion from materials outside
$a_{\rm p}$ well into the gap region.  After 2440 orbits, the tracers' density $\Sigma_{\rm d}$ is nearly uniform.  Relatively large viscosity also
enhance angular momentum transfer and mass diffusion across the interface between horseshoe streamlines and CPD.  Consequently
the vortensity gradient is smoothed out such that the obstruction of gas flow across the tidal barrier and onto the CPD is 
reduced.  These effects elevate the mass supply onto the CPD and eventually accretion into the sink hole.  

Similar flow pattern of the gas have also been noted by \citet{Lubowetal1999} and \citet{Miyoshi1999}.  We reiterate
\citet{Dobbsdixon2007}'s conclusions that the physical meaning behind the separation of flows is the mismatch of vortensity. 
The streamline/velocity plots for the tracer particles that are added to couple with the gas \textit{after} the formation of the gap provide more
clear illustration on the nature of the accretion flows, especially for low-viscosity cases, by avoiding the interference of horseshoe flow around $L_4$ \& $L_5$ points that's irrelevant to incoming materials.

\section{Discussions}
\label{sec:discussion}
\subsection{2D and 3D simulations}
\label{2d3d}
{We have used global 2D high-resolution simulations in this study. For gap opening planets with $R_{\rm H}>H_{\rm p}$, this approximation captures most of the the vertically 
averaged flow pattern of the gas around
the planet, but still it may have excluded some important physical effects. \citet{Szulagyi2014} finds that certain mechanisms in an 
inviscid 3D CPD structure can also quench planet growth, when the planet is actually fed by meridional flows that transport materials to its poles \citep{FungChiang2016}. Radiative transfer and different equations of state are also worth considering \citep{Aycliffe2009,Szulagyi2016}. }

{In particular, results from 3D nested-grid simulations of \citet{Boden2013}
do not show significant rise of accretion rate for $M_{\rm p}\gtrsim 5\ M_{\rm J}$ resulting from unstable streamline eccentricity.
Their results contradict the 
conclusions of many global 2D simulations \citep{KleyDirksen2006,DuffelChiang2015,TeysOgilve2017, zhang2018} 
as well as our findings. { Yet, as discussed below, our 3D simulations also point toward a dichotomy, possibly associated to the 2D vs 3D geometry of the system.}}

\begin{figure}[htbp]
\centering
\includegraphics[width=0.45\textwidth,clip=true]{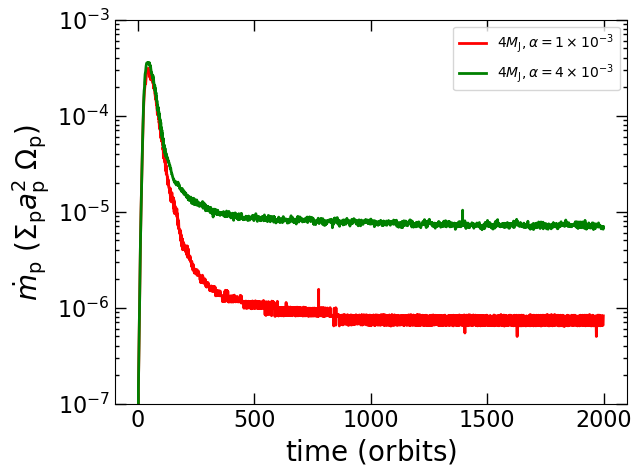}
\includegraphics[width=0.45\textwidth,clip=true]{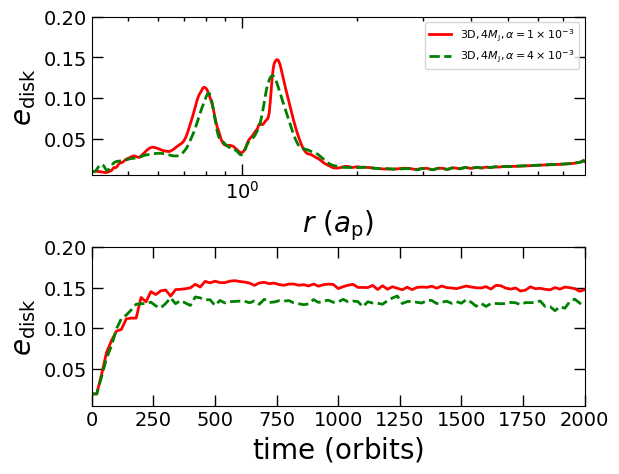}
\caption{Planetary accretion rate and disk eccentricity for the 3D simulation with a planet mass of $4\ M_{\rm J}$, $h_{\rm p}=0.05$. The red lines correspond to $\alpha=1\times10^{-3}$, while the green lines are for $\alpha=4\times10^{-3}$. The other disk parameters are the same the 2D runs. Top panel: the time evolution of the planetary accretion rate. Middle panel: The azimuthally averaged radial distribution of disk eccentricity at $t=2000$ orbits. Lower panel: the growth of the \textit{peak} eccentricity within the disk over time. }
 \label{fig:planet3d}
\end{figure}

%

{Could low resolution for gas far from the planet in the nested-grid coordinates account for this discrepancy? Comparing the different resolution runs in \citet{KleyDirksen2006}, and our results with a finer resolution, it appears that resolution does not intrinsically affect eccentricity excitation, at least in 2D. Another potential caveat is the possibility that the 3D simulations
with the nested grid were performed with insufficient amount of time to allow for eccentricity modes to grow. In order to
find some clues to the cause of this dichotomy, we perform additional high resolution 3D runs as listed in Table \ref{tab:para}.}

{Instead of measuring the accretion rate after a few hundred orbits as it was done in these previous 3D simulations, 
our calculations were followed for two thousand orbits.  This duration is much longer than the growth time of potential 
eccentric modes in the 2D simulations. We plot our results in Figure \ref{fig:planet3d}, 
in a way that is analogous to Figures \ref{fig:eccentric} and \ref{fig:eccentricevolution}.}

{In the first 3D simulation (red solid lines in Figure \ref{fig:planet3d}), we adopt $h_{\rm p}=0.05$ 
and $\alpha=10^{-3}$.  For this set of disk parameters, a planet with $M_p \sim 2M_{\rm J}$ would be
sufficiently massive to excite unstable streamline
eccentricities in 2D limit.  In contrast, the peak eccentricity in 3D is quite 
stable for $M_p= 4M_{\rm J}$, and the accretion rate does not rise significantly even after 2000 orbits. We have also plotted the final accretion rate of this run in the upper right panel of Figure \ref{summary} with similar parameter space. In another run
(green solid lines), we use the same viscosity parameter $
\alpha=4\times 10^{-3}$ at $a_{\rm p}$ as \citet{Boden2013}. We obtain a low steady accretion rate that 
corresponds to $4\times 10^{-6} \Sigma_{\rm p}a_{\rm p}^2\ {\rm yr^{-1}}$ in the units of Figure 
\ref{fig:predictions}, which is a factor of 2 smaller than the result of \citet{Boden2013} for $4M_{\rm J}$, 
with no intrinsic difference caused by eccentricity.}

{Our 3D results suggest that resolution and limited simulation time may not be the culprit of eccentricity damping. If anything, 
we note that higher resolution seem to slightly reduce the orderly-accretion rate, rather than cause significant rise of it due to 
eccentricity excitation. It is possible that there are intrinsic differences between how gas eccentricity is excited under the 2D
versus 3D topography. Whether even higher uniform resolution could change this result or not warrants further investigation. But 
3D simulations with even higher resolution and higher planet mass require much more computational resources and they are also
subject to additional technical challenges. Nevertheless, the polar accretion effect and eccentricity damping in 3D flow, if they 
are indeed due to physical rather than numerical artefacts, would both disfavor high ${\dot m}_{\rm p}$ and contribute to quenching the accretion 
rate.}

\subsection{The Inviscid Regime}

\begin{figure}[htbp]
\centering
\includegraphics[width=0.45\textwidth,clip=true]{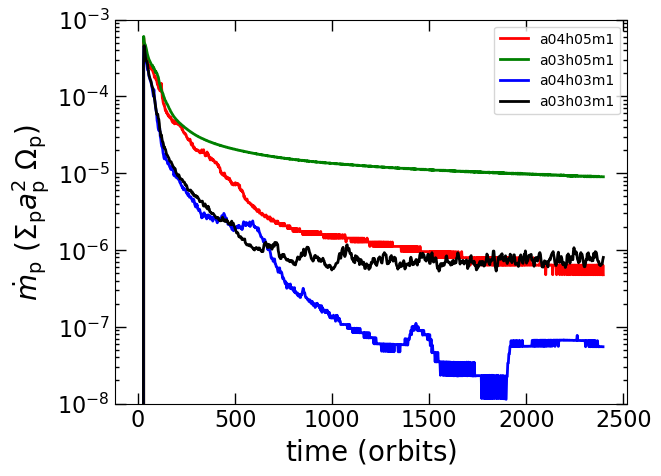}
\includegraphics[width=0.45\textwidth,clip=true]{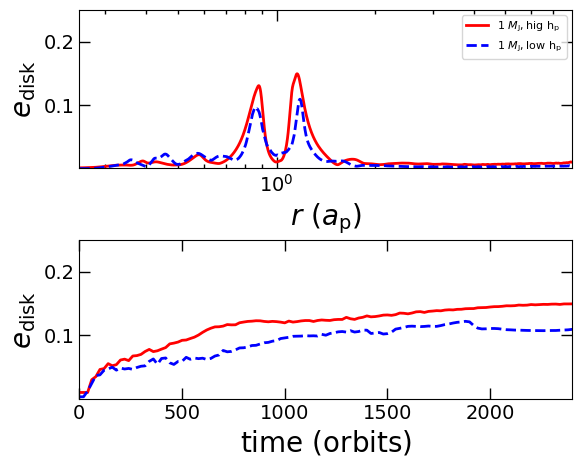}
\caption{Planetary accretion rate and disk eccentricity with a very low disk viscosity simulations. Top panel: the time evolution of the planetary accretion rate. The red and blue lines in the upper panels correspond to $\alpha=1.11\times10^{-4}$, while the green and black lines are for $\alpha=1.11\times10^{-3}$. The other disk parameters are indicated by the model names.  Middle panel: The azimuthally averaged radial distribution of disk eccentricity at $t=2400$ orbits for two lowest viscosity $\alpha=1.11\times10^{-4}$ models. Lower panel: the growth of the \textit{peak} eccentricity within the disk over time for two $\alpha=1.11\times10^{-4}$ models. Different colors in the middle and lower panels represent different $h_{\rm p}$ as in the upper panel.}
 \label{fig:planet_invis}
\end{figure}

{Recent models of ALMA maps of protostellar disks suggest that weak turbulence $\alpha \leq 10^{-3}$ may be 
a common feature of planet-forming disks \citep{zhang2018, Flaherty2020}. Therefore, it is also meaningful to probe 
even lower viscosity regime with high resolution. In Figure \ref{fig:planet_invis}, we plot results from additional 
2D simulations for the very low viscosity regime. The top panel shows the evolution of ${\dot m}_{\rm p}$ for the
simulations with low and moderate viscosity. } 

{In model \texttt{a04h05m1} (red solid lines), all parameters are the same as in model \texttt{a03h05m1} (green solid line in the top panel) except that $\alpha =1.11\times 10^{-4}$. As expected by \citet{Dobbsdixon2007}, the tidal barrier is more effective and the accretion rate is further suppressed.}

{All the parameters in model \texttt{a04h03m1} (blue solid lines) are the same as those in model \texttt{a03h03m1} 
(black solid line in the top panel) except the viscosity is lower by a factor of ten. In this case, the peak eccentricity 
fluctuated slightly between $t=1500-2000$ orbits.  It is accompanied by a mild rise in accretion rate, before it settles 
down to a steady value. This pattern is not surprising since model \texttt{a03h03m1} is already on the verge of instability 
(doubling in planet mass would lead in unstable streamline eccentricity).  With a lower viscosity, the reduced diffusion 
rate decreases the damping efficiency for the excited streamline eccentricities. Nevertheless, planet's asymptotic accretion 
rate is much smaller compared with that in model \texttt{a03h03m1}.  This result reinforces the conclusion that the planet's 
mass can be very robustly retained at $\sim M_{\rm J}$ in the low-viscosity limit.}

\subsection{Accretion of a Migrating Planet}\label{sec:mig}

\begin{figure}[htbp]
\centering
\includegraphics[width=0.45\textwidth,clip=true]{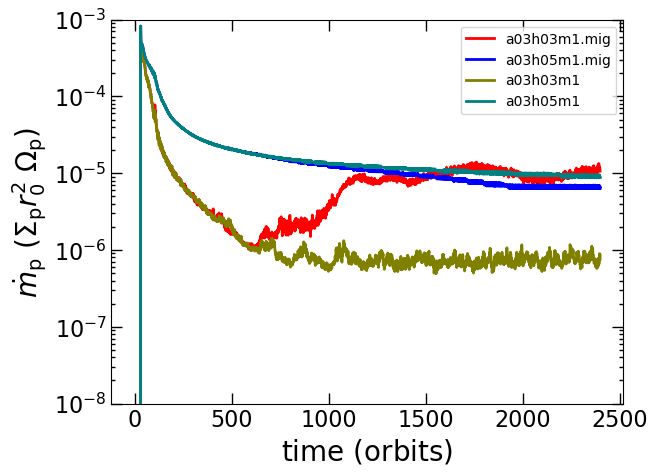}
\includegraphics[width=0.45\textwidth,clip=true]{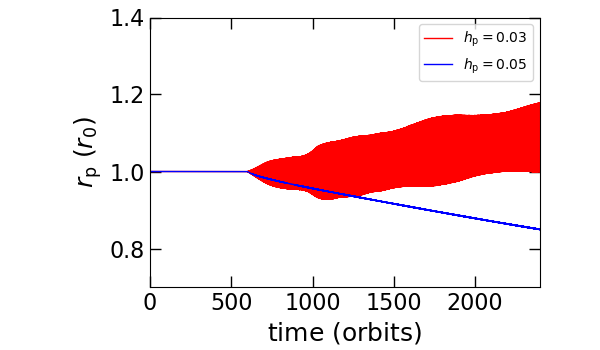}
\caption{Planetary accretion rate and orbital evolution for migrating planet models. Top panel: the time evolution of the planetary accretion rate measured in units of $\Sigma_{\rm p} r_0^2\Omega_{\rm p}$. Here $r_{0}=1$ is fixed to be the initial orbital radius of the planet, and $\Omega_{\rm p},\Sigma_{\rm p}$ are evaluated at $r_0$ and $t=0$. The red and blue lines in the upper panels correspond to two migrating planet models, while the other two lines are the corresponding non-migrating planet cases. The other disk parameters are indicated by the model names.  Lower panel: the orbital evolution of two migrating planet cases. The low $h_{\rm p}$ case shows a strong planetary eccentricity excitation.}
 \label{fig:planet_mig}
\end{figure}

{In our numerical simulations, we neglect the feedback torque of the disk gas on the planet's orbit.
Non vanishing torque leads to angular momentum exchange between the planet and its natal disk.  The resulting
planetary migration may, in turn,  affect the diffusion across the gap/CPD and accretion rate of the planet
\citep{nelson2000,DAngeloetal2005,DAngelo2008}. In a recent paper, \citet{Durmannkley2017} used uniform grid 
2D simulation to investigate the accretion rate of migrating planets. They impose a large sinkhole (with
$\Delta = 0.5R_{\rm H}$) to represent the embedded planet in their simulation.
They experimented with different surface density $\Sigma_{\rm min}$ and accretion efficiency $f/\tau$ to 
obtain a range of accretion rates, although their sinkhole is too large for the results to converge.
Nevertheless, they reach the conclusion that the removal of gas close to the planet reduces the feedback
torques and slows down a migration rate. Here we perform some additional simulations for a migrating planet 
to directly compare with the non-migrating cases. We compare the accretion rate evolution of these cases 
in the top panel of Figure 
\ref{fig:planet_mig}, and plot the evolution tracks of its distance to the star $r_{\rm p}$ in the 
lower panel.}

{In model \texttt{a03h05m1.mig} (blue solid lines), all the parameters are identical to those in model 
\texttt{a03h03m1} except we release the planet at $t=600$ orbits and let it move freely under the influence 
of disk torques. We find that the accretion rate is not affected by the release, albeit there is considerable 
inward migration. In model \texttt{a03h03m1.mig} (red solid lines), however, eccentricity of the streamlines 
and that of the planet's orbit are excited. We have commented before that case \texttt{a03h03m1} is already 
on the verge of eccentricity excitation.  The additional degree of freedom provided by the planet's release 
is likely to break the stability and significantly enhances the accretion rate. }

{This result suggests that realistic migration might change our results for some cases in which
eccentric streamlines are marginally stable (Fig. \ref{summary}). On the other hand, various mechanisms which may halt migration has not been considered in this study.  These
effects include dust feedback \citep{Kanagawa2019}, or a steeper initial surface density slope of disk gas 
$s$ \citep{chen2020b}. In subsequent studies, we will explore a larger parameter space of disk conditions 
to self-consistently study both accretion \& migration of the planets coherently.}

\subsection{Summary}
\label{sum}
In this paper, we performed extensive numerical simulations to study the dynamical accretion of gas giants. Some previous works
\citep{Tanigawa2016,rosenthal2020consumption} have estimated the asymptotic masses of gas giants, with empirical \& semi-analytical TT and RCGM 
prescription. These formulae predict that the accretion rates falls off as power-law functions of $M_{\rm p}$ 
after the planets have become sufficiently massive to open up gaps in their natal disks near their orbital semi major axis. 
In the determination of ${\dot m}_{\rm p}$, the tidal perturbation of the host star inside the planets' Hills radius is generally 
neglected. Moreover, the background is assumed to be provided by the residual gas along the horseshoe streamlines with
the azimuthally averaged lowest residual surface density found from numerical simulation and analytic modeling \citep{DM13,Kanagawaetal2015MNRAS}.
Under these assumptions, only deep gap-clearing or severe global depletion could quench the accretion onto the planet. 
In disks with moderate or large viscosity, these prescriptions predict {typical} asymptotic masses $\gtrsim 10 M_{\rm J}$
which is difficult to reconcile with observational data. In addition, the mass ratio of such planets to their host
stars $q (\gtrsim 0.005)$ exceeds the critical values which can lead to the excitation of unstable streamline eccentricity
\citep[e.g.][]{KleyDirksen2006,DuffelChiang2015} and further elevation of planets' accretion rate and asymptotic masses in 2D scenarios.
 
In contrast, \citet{Dobbsdixon2007} used analytical approximations and 2D numerical simulations to 
show that, in the low-viscosity or inviscid limit, 
the conservation of Bernoulli constant and vortensity of gas streamlines around the planet contribute together
to impose a tidal barrier.  Inside the planet's Hill sphere, the gas streamlines are deflected by the host
stars' tidal perturbation and the Coriolis force from the frame which corotates with the orbit of the planet.
The tidal barrier separates the incoming horseshoe streamlines from those on the CPD. Across
the barrier, there is a vortensity discontinuity. Viscous stress due to either intrinsic viscosity or shock
dissipation leads to angular momentum transfer, and the diffusion of a small fraction of gas on the incoming
horseshoe streamlines onto the outer regions of the CPD.  This continuous influx of gas is eventually accreted 
by the planet as it viscously diffuse through the CPD.  
 
\citet{Dobbsdixon2007} has numerically verified this theory in inviscid disks.  They also introduced a DLL 
prescription for the planets' asymptotic mass.
We carry out a series of models 
with a range of aspect ratio and viscosity.  The relevance of $h_{\rm p}$ is clearly stated in Equation \ref{eq:abarrier}.
The magnitude of $\alpha$ is also important in regulating the rate of angular momentum transfer and mass diffusion
1) between the horseshoe streamlines and PSD outside (and inside) the gap and 2) across the tidal barrier between
horseshoe streamlines and CPDs.  

In Table \ref{tab:para} we summarize the simulation parameters and the normalized planetary accretion rates for 
all of our models. We compare the numerical results with various prescriptions.  In low-viscosity disks, the relation 
between stable-eccentricity accretion rate and planet mass conform more with exponential decay as predicted by 
the DLL prescription than the linear scalings from the TT and RCGM prescription. For high-viscosity disks, the 
tidal barrier is overcome by diffusion and dissipation such that the decline in ${\dot m}_{\rm p}$ for
increasing $M_{\rm p}$ is less steep. 
We used vortensity contours and tracer velocity arrowplots to show that the ``barrier" deflecting accreting streamlines is directly 
associated with the distribution of vortensity. Tanaka, Yuki., et al. (in prep) are performing a parallel 
numerical study on orderly \& eccentric accretion of gas giants, nevertheless their focus is different from the tidal barrier.

When appropriate scaling for physical parameters is applied, overgrown gas giants (with $M_{\rm p} \gg M_{\rm J}$) can be 
avoided in PSDs around classical T Tauri stars provided their $\alpha \sim 10^{-3}$ and $h_{\rm p}
\sim 0.03-0.05$.  In this low-viscosity limit, the tidal barrier model is an appropriate prescription for the 
asymptotic mass of planets.  In highly viscous disks (with $\alpha \gtrsim 10^{-2}$), the tidal barrier is no longer 
effective and the asymptotic mass of planets formed in orderly disks around classical T Tauri stars are typically 
several times more massive than Jupiter.  Similarly, thicker disks imply larger $q_{\rm th}$ and super Jupiter
masses. Such massive planets may excite unstable streamline
eccentricity, enhance angular momentum transfer and mass supply to the outer regions of CPD
and further elevate the asymptotic mass of gas giant planets to more than an order of magnitude larger than that
of Jupiter.  Based on these inference and the observed mass distribution of exoplanets, we speculate that
most Jupiter-mass planets are formed in PSDs with relatively low $\alpha (\sim 10^{-3}$). 

{{In the discussions, we perform additional separate simulations for 3D geometry, inviscid disks and migrating planets. 
The results suggest that: i) the absence of eccentricity excitation in 3D are more likely to be associated with the intrinsic difference between how eccentricity is excited in the 
two geometries, rather than resolution or simulation 
timescale issues; ii) in disks with very low viscosity ($\alpha \sim 10^{-4}$), the tidal barrier is more effective in the orderly-accreting cases (in marginally stable cases the streamline eccentricity may be briefly excited, but 
the accretion rate is still severely suppressed); iii) planet migration might significantly boost the accretion rate by 
exciting eccentricity in the marginal cases. More disk parameters and physical processes needs to be considered in 
subsequent high resolution simulations to coherently study planet migration and accretion.}}

 
For the theoretical predictions in Figures \ref{fig:predictions} and \ref{summary}, we have extended the \citet{DM13,Kanagawaetal2015MNRAS} gap-scaling of $\Sigma_{\rm min}$ to at least $4\ M_{\rm J}$ to compare with our 
numerical results, as in \citet{Tanigawa2016}. There is no physical reason to justify the extrapolation of this 
scaling law beyond the gap opening limit since the Lindblad torques at the edges of the gap no longer uniformly scales
with the bottom density. Empirically, it can still be used to inferred the average torque density winthin an order unity 
provided there is still some residual gas flowing across the gap \citep{chen2020b}. In 3D simulations, this gas flow is never completely cut off \citep[e.g.][]{Szulagyi2014}. A more accurate analytical surface
density formulae for gap-opening planets needs to be constructed in order to reduce this uncertainty in the future. We note that 
if this gap scaling law overestimates $\Sigma_{\rm min}$ for the gap-opening planets, the planetary 
accretion rates would be smaller.  It would also imply smaller (and more consistent with observation) asymptotic masses 
for gas giants. Although we have shown numerically that the effectiveness of the tidal barrier is diminished 
by viscous dissipation, an analytical model on how tidal barrier can be partially overcome in the presence 
of moderate viscosity is left for future theoretical modeling.

\acknowledgements
We thank Eugene Chiang, Chris Ormel, Gordon Ogilvie, Hidekazu Tanaka, Hui Li, and Judit Szulagyi for helpful discussions. We thank Yuki Tanaka for sharing information about their paper in preparation. We thank the anonymous referees for providing suggestions that significantly improved this work. Y.P.L. gratefully acknowledges the support by LANL/LDRD. 
This research used resources provided by the Los Alamos National Laboratory Institutional Computing Program, which is supported by the U.S. Department of Energy National Nuclear Security Administration under Contract No. 89233218CNA000001. 
Softwares: \texttt{LA-COMPASS} \citep{Lietal2005,Lietal2009}, \texttt{Numpy} \citep{2011CSE....13b..22V}, \texttt{Matplotlib} \citep{2007CSE.....9...90H}

\bibliography{revision}
\bibliographystyle{aasjournal}

\end{document}